\documentclass[letterpaper,titlepage,12pt]{article}
\usepackage[colorlinks=true,urlcolor=blue,citecolor=magenta]{hyperref}
\usepackage{amssymb,amsmath,amsfonts}
\usepackage{epsfig}
\usepackage{epsfig}
\usepackage{graphicx}
\usepackage{epstopdf}
\usepackage{caption}
\usepackage{subcaption}
\usepackage{amscd}
\usepackage{amsthm}
\usepackage{latexsym}
\usepackage{amsbsy}
\usepackage{bbm}
\usepackage[english]{babel}
\usepackage{psfrag}
\usepackage{tabularx}
\usepackage{babel}
\usepackage{varioref}
\usepackage{amsmath}
\usepackage{amsthm}
\usepackage{amssymb}
\usepackage{esint}
\usepackage {hyperref}
\usepackage{breakurl}

\allowdisplaybreaks

\def\clock{{\count0=\time
           \divide\count0 60
           \ifnum\count0<10 0\fi\the\count0
           \multiply\count0 -60 \advance\count0 \time
           :\ifnum\count0<10 0\fi \the\count0
         }}
\newcommand{\timestamp}{{\small\vbox{\hbox{\tt\jobname.tex}
\hbox{\the\day/\the\month/\the\year, \clock}}}}

\numberwithin{equation}{section}

\begin{document}

\begin{titlepage}
\rightline{\vbox{   \phantom{ghost} }}
%
%
 \vskip 1.4 cm
\centerline{\LARGE \bf Torsional Newton-Cartan Geometry 
}
\vspace{.2cm}
\centerline{\LARGE \bf  from the Noether Procedure}

\vskip 1.5cm

\centerline{ {\bf  Guido Festuccia$^1$,  Dennis Hansen$^2$, Jelle Hartong$^3$, Niels A. Obers$^2$}}

\vskip .8cm

\begin{center}

\sl $^1$ Department of Physics and Astronomy, Uppsala University, \\ 
\sl  SE-751 08 Uppsala, Sweden \\
\sl $^2$ The Niels Bohr Institute, Copenhagen University,\\
\sl  Blegdamsvej 17, DK-2100 Copenhagen \O , Denmark. \\ 
\sl $^3$ Physique Th\'eorique et Math\'ematique and International Solvay Institutes,\\
Universit\'e Libre de Bruxelles, C.P. 231, 1050 Brussels, Belgium

\vskip 0.4cm

\end{center}
\vskip 0.6cm


\vskip .8cm \centerline{\bf Abstract} \vskip 0.2cm \noindent

We apply the Noether procedure for gauging space-time symmetries to theories with Galilean symmetries, analyzing both massless and massive (Bargmann) realizations. It is shown that at the linearized level the Noether procedure gives rise to (linearized) torsional Newton--Cartan geometry. In the case of Bargmann theories  the Newton--Cartan form $M_\mu$ couples to the conserved mass current. We show that even in the case of theories with massless Galilean symmetries it is necessary to introduce the form $M_\mu$ and that it couples to a topological current. Further, we show that the Noether procedure naturally gives rise to a distinguished affine (Christoffel type) connection that is linear in $M_\mu$ and torsionful. As an application of these techniques we study the coupling of Galilean electrodynamics to TNC geometry at the linearized level.

\end{titlepage}

\tableofcontents

\section{Introduction}

Field theories are central to the description of a wide range of physical phenomena. Their understanding continues to be an important theme, in which  a crucial role is played by both spacetime and internal symmetries. One way of classifying space-time symmetries is in terms of their subset of boost symmetries, 
which come in three types,  Lorentz, Galilean and Carrollian, corresponding to relativistic, non-relativistic and ultra-relativistic boost symmetries respectively. 
In each of these cases the theory can also be made scale invariant, while symmetry breaking patterns of different kinds can arise as well.  
A  natural question to ask is how to formulate the coupling of such classes of field theories to relevant geometric structures while respecting diffeomorphism invariance. One way to address this question at the linearized level is to consider the Noether procedure and gauge the space-time symmetries. Finding the resulting underlying  geometry opens up for a host of interesting applications, including the study
of Ward identities, anomalies, hydrodynamics, holographic realizations and new dynamical theories of gravity. 

In this context, Newton-Cartan (NC) geometry \cite{Cartan1,Cartan2,Friedrichs1928}
has in recent years seen an increasing interest, in part 
due to its appearance as the boundary geometry
 \cite{Christensen:2013lma,Christensen:2013rfa}
in holographic setups with anisotropic scaling in the bulk \cite{Son:2008ye,Balasubramanian:2008dm,Kachru:2008yh,Taylor:2008tg}.
More generally this has been motivated, as explained above, from a wider  field theoretic perspective as the background geometry to which non-relativistic field theories couple in a covariant way. In particular, this followed the proposal in  \cite{Son:2013rqa}  to use NC geometry in constructing
an effective field theory for quantum Hall states.  Furthermore, NC geometry and its torsionful generalization appears to be the 
natural geometrical framework underlying Ho\v rava-Lifshitz gravity, allowing for full diffeomorphism invariance \cite{Hartong:2015zia}.

A generalization of NC geometry with torsion, called torsional Newton-Cartan (TNC) geometry, was  first observed  \cite{Christensen:2013lma,Christensen:2013rfa}
in the context of holography for a  specific action supporting $z=2$ Lifshitz geometries.
This analysis was subsequently generalized to a large class of holographic Lifshitz models for arbitrary values of $z$  in \cite{Hartong:2014oma,Hartong:2015wxa}. 
In parallel, TNC geometry was shown to arise from gauging the  Schr\"odinger algebra \cite{Bergshoeff:2014uea} or the Bargmann algebra \cite{Hartong:2015zia},
generalizing earlier work \cite{Andringa:2010it} that showed how to obtain NC geometry from gauging Bargmann. 
In applications to condensed matter systems  with non-relativistic symmetries,  such as the fractional quantum Hall effect, 
 the addition of (twistless) torsion was presented in  \cite{Geracie:2014nka} following the introduction  \cite{Son:2013rqa} of NC geometry to this problem. 
 The coupling of non-relativistic field theories to TNC was also considered in 
  \cite{Jensen:2014aia,Hartong:2014pma,Hartong:2015wxa}. 
Further investigations of TNC geometry from different perspectives subsequently appeared in 
 \cite{Bekaert:2014bwa,Geracie:2015dea,Bekaert:2015xua} and \cite{Banerjee:2014nja,Brauner:2014jaa,Banerjee:2016laq}.

The relevant geometric fields in TNC are a time-like vielbein $\tau_\mu$ and inverse spatial metric  $h^{\mu \nu}$ with corank 1 plus
their projective inverses, together with a one-form $M_\mu$.%
\footnote{In cases where there is no particle number $U(1)$ symmetry, it is useful to introduce a St\"uckelberg scalar $\chi$
and write $M_\mu=m_\mu-\partial_\mu\chi$ where $m_\mu$ transforms as a connection under $U(1)$ particle number (see e.g.  \cite{Christensen:2013lma,Hartong:2015wxa,Hartong:2015zia}).
With the $\chi$-field one can also couple non-relativistic theories that do not have a local $U(1)$ symmetry
to TNC geometry. In this paper we will restrict to the case of field theories that have at least Galilean symmetries.}
One of the features that distinguishes TNC geometry from Riemannian geometry is that, while in the latter there is a unique metric compatible and torsionless affine
connection (the Levi-Civita connection),  in TNC geometry this is not the case.
Furthermore, for all Galilean affine connections the temporal part of the torsion is completely fixed and proportional to $\partial_\mu\tau_\nu-\partial_\nu\tau_\mu$ \cite{Bekaert:2014bwa,Bekaert:2015xua}.
Thus, torsion appears quite naturally and generally  if one does not require any conditions on the flow of time $\tau_\mu$.
The original case of NC geometry%
\footnote{See also \cite{Duval:2011mi,Duval:2016tzi} for recent work.}
 assumes that $\tau_\mu$ is closed, i.e. providing a notion of absolute time, making torsionlessness possible.%
\footnote{The in-between case in which $\tau_\mu$ is hypersurface orthogonal is called twistless torsional NC (TTNC) geometry.}
Already in this case, one finds that there is no unique  connection that solves the analogue
of metric compatibility and the torsionless condition. Such Galilean connections are only determined up to an arbitrary two-form due to the degeneracy of $h^{\mu \nu}$ \cite{Bekaert:2014bwa,Bekaert:2015xua}. Thus, contrary to Riemannian geometry
where zero torsion has the advantage of selecting a unique distinguished connection, for NC geometry this is not the case.
Moreover, there is another field-theoretic reason why torsion is natural to consider in the non-relativistic case. This is because
energy density and energy flux are the response to varying $ \tau_\mu$, so that in order to be able to compute the most general response
this quantity better be unconstrained.

Thus, it seems that there is no distinguished connection in (T)NC geometry, and, moreover, it appears that the various approaches in the literature lead to different parametrizations of the connections.  A natural question to ask is thus, whether there exists a \textit{minimal connection} 
in the sense that it requires the least number of gauge fields in order to define a good covariant derivative.
For Riemannian geometry the minimal connection is identical to the  Levi-Civita connection, which can be expressed in terms of just the vielbeins, but for TNC geometry the situation is more complicated.

The purpose of this paper is to employ the Noether procedure to the gauging of space-time symmetries in  non-relativistic field theories  in order to determine the minimal connection of TNC geometry.
To this end, we will consider  field theories that have  Galilean and Bargmann spacetime symmetries, depending
on whether they are massless or massive respectively \cite{ raey52,Bargmann:1954gh,LevyLeblond:1967zz,LevyLeblond1971}. 
Interestingly, this minimal connection turns out to be the one that was identified 
 among the one-parameter family of \cite{Hartong:2015zia}, as the  unique connection that satisfies the additional requirement that the connection is linear in $M_\mu$. 
Our treatment can, moreover, be seen as a completely general field-theoretic construction of TNC geometry at the linear level
(see e.g.  \cite{VanNieuwenhuizen:1981ae,Kraus:1992gk,Ortin2004} for this construction in the context of general relativity).

 A brief outline of the paper is as follows. 
In section \ref{TNC_geom} we review some of the most important aspects of TNC and Galilean connections.
We will then use the Noether procedure to analyze Galilean and Bargmann cases separately in sections \ref{noether_gal} and \ref{noether_barg}.

In section \ref{sec:Comparison to usual procedure} we compare the results to the full non-linear TNC connection. Finally we study Galilean Electrodynamics (GED) as a concrete example in section \ref{Galdelectroynamics}. A much more complete analysis
of  GED will be presented in the companion paper \cite{Festuccia:2016caf}.
We perform in appendix \ref{sec:The-Noether-procedure-lorentz} the analogous construction for relativistic theories, 
to compare to this more familiar case. Finally, appendix \ref{sec:lin_NC} contains some details on the linearization of
TNC geometry around flat NC spacetime.

\section{Torsional Newton-Cartan geometry} \label{TNC_geom}

This section is designed to provide a brief overview of Newton--Cartan geometry to set the stage for the later sections on the Noether procedure and, in particular, to enable a direct comparison between the geometrical approach reviewed here and the field theory approach discussed in sections \ref{noether_gal} and \ref{noether_barg}. Most of this material  is taken from  \cite{Hartong:2015zia}.

\subsection{Gauging the Galilean and Bargmann groups}\label{TNC_gauge}

The symmetry algebra of Galilean theories
consists of generators of time translation $H$, spatial translations $P_{i}$,
Galilean boosts $G_{i}$ and spatial rotations $J_{ij}$. The defining
non-zero commutation relations are given by

\begin{subequations}\label{eq:Galilean_algebra}

\begin{eqnarray}
\left[J_{ij},P_{k}\right] & = & -\delta_{ik}P_{j}+\delta_{jk}P_{i} \label{eq:rotation_momentum_comm}\\
\left[J_{ij},J_{kl}\right] & = & \delta_{ik}J_{jl}-\delta_{il}J_{jk}-\delta_{jk}J_{il}+\delta_{jl}J_{ik}\\
\left[H,G_{i}\right] & = & -P_{i} \label{eq:hamilton_boost_comm}\\
\left[J_{ij},G_{k}\right] & = & -\delta_{ik}G_{j}+\delta_{jk}G_{i}\,.
\end{eqnarray}

\end{subequations}

This algebra has a double direct sum structure, where the boosts form
a direct sum with the rotations, and translations form a direct sum
with both of these. Moreover translations and boosts give rise to two abelian subalgebras. The boosts and momenta transform
as vectors under rotations. The latter form an $\mathfrak{so}\left(d\right)$
subalgebra. The structure of the $D=d+1$ dimensional Galilean group
$\mathrm{Gal}\left(d,1\right)$ can thus be summarized as
\begin{equation}
\mathrm{Gal}\left(d,1\right)=\mathbb{R}^{d+1}\ltimes\left(\mathbb{R}^{d}\ltimes\mathrm{SO}\left(d\right)\right)\,.\label{eq:Galilean_group}
\end{equation}

The Galilean group has a central extension in any dimension known as the Bargmann group \cite{Bargmann:1954gh}.
The group structure that extends \eqref{eq:Galilean_group} is given by
\begin{equation}
\mathrm{Barg}\left(d,1\right)=\left(\mathbb{R}\ltimes\mathbb{R}^{d+1}\right)\ltimes\left(\mathbb{R}^{d}\ltimes\mathrm{SO}\left(d\right)\right)\,. \label{eq:Bargmann_group}
\end{equation}
The  commutator $\left[P_{i},G_{j}\right]$ is no longer vanishing 
\begin{equation} \label{eq:Bargmann_algebra}
\left[P_{i},G_{j}\right]=N\delta_{ij}\,.
\end{equation}
Here $N$ is the central element called the mass or particle number generator. For more details see for example \cite{Duval:1984cj,Hagen:1972pd}.

\begin{table}
\centering{}%
\begin{tabular}{|c||c|c|c|c|}
\hline 
Symmetry & Generator & Gauge field & Gauge parameter & Curvature\tabularnewline
\hline 
\hline 
Time translations & $H$ & $\tau_{\mu}$ & $\tau_{\mu}\xi^{\mu}$ & $R_{\mu\nu}\left(H\right)$\tabularnewline
\hline 
Space translations & $P_{i}$ & $e_{\mu}^{i}$ & $e_{\mu}^{i}\xi^{\mu}$ & $R_{\mu\nu}{}^{i}\left(P\right)$\tabularnewline
\hline 
Galilean boost & $G_{i}$ & $\Omega_{\mu}{}^{i}$ & $\tilde{\lambda}^{i}$ & $R_{\mu\nu}{}^{i}\left(G\right)$\tabularnewline
\hline 
Spatial rotations & $J_{ij}$ & $\Omega_{\mu}{}^{ij}$ & $\tilde{\lambda}^{ij}$ & $R_{\mu\nu}{}^{ij}\left(J\right)$\tabularnewline
\hline 
Central & $N$ & $M_{\mu}$ & $\sigma$ & $R_{\mu\nu}\left(N\right)$\tabularnewline
\hline 
\end{tabular}\caption{Gauge fields and parameters in the gauging of the Galilean and Bargmann groups.\label{table_gauging}}
\end{table}

The goal of this paper is to derive torsional Newton--Cartan (TNC) geometry from the Noether procedure applied to Galilean and Bargmann invariant field theories. However, before going into that we first summarize previous work on TNC geometry that is based on a more geometrical approach following \cite{Andringa:2010it,Jensen:2014aia,Hartong:2014pma,Bergshoeff:2014uea,Hartong:2015zia}. This involves manifolds whose tangent space symmetries are dictated by the $G_i$, $J_{ij}$ and $N$ generators of the Bargmann algebra while the general coordinate transformations will be obtained by a deformation of the transformations generated by $H$ and $P_i$.

In this geometrical approach we gauge the Bargmann algebra \eqref{eq:Galilean_algebra}, \eqref{eq:Bargmann_algebra} 
by introducing gauge fields corresponding to generators as in table \ref{table_gauging}. Let us introduce the Yang--Mills connection $A_\mu$ and its field strength $F_{\mu\nu}$ as\footnote{ Here and in the following we denote antisymmterization over indices with $[]$ and symmetrization with $()$. For instance $T_{[i_1,\ldots i_n]}={1\over n!} \sum_{\sigma\in S_n} (-1)^\sigma T_{i_{\sigma(1)}\ldots i_{\sigma(n)}}$.}
\begin{subequations}
\begin{eqnarray}
A_{\mu} & = & H\tau_{\mu}+P_{i}e_{\mu}^{i}+G_{i}\Omega_{\mu}{}^{i}+\frac{1}{2}J_{ij}\Omega_{\mu}{}^{ij}+NM_{\mu}\,,\\
F_{\mu\nu} & = & 2\partial_{[\mu}A_{\nu]}+\left[A_{\mu},A_{\nu}\right]\nonumber\\
 & = & HR_{\mu\nu}\left(H\right)+P_{i}R_{\mu\nu}{}^{i}\left(P\right)+G_{i}R_{\mu\nu}{}^{i}\left(G\right) +\frac{1}{2}J_{ij}R_{\mu\nu}{}^{ij}\left(J\right)+NR_{\mu\nu}\left(N\right)\,.
\end{eqnarray}
\end{subequations}
The gauge field transforms in the adjoint 
\begin{equation}
\delta A_{\mu} = \partial_{\mu}\Lambda+\left[A_{\mu},\Lambda\right]\,.
\end{equation}

The local algebra of transformations can be deformed in such a way that the local translations of the gauge transformation become the generators of infinitesimal general coordinate transformations (GCTs). This is achieved by setting $\Lambda=\xi^\mu A_\mu+\Sigma$ where $\Sigma$ does not contain the $H$ and $P_i$ generators and by
defining a new transformation as
\begin{equation}
\overline{\delta} A_{\mu}=\delta A_{\mu}-\xi^{\nu}F_{\mu\nu}=\mathcal{L}_{\boldsymbol{\xi}}A_{\mu}+\partial_{\mu}\Sigma+\left[A_{\mu},\Sigma\right]\,.
\end{equation}
The $\Sigma$ transformations correspond to local tangent space transformations.

The connections $e_\mu^i$ and $\tau_\mu$ are now interpreted as vielbeins and we can thus introduce inverse vielbeins  $v^\mu, e^\mu_i$ by requiring that they satisfy
\begin{equation}
v^\mu\tau_\mu=-1\,,\qquad v^\mu e_\mu^i=0\,,\qquad e^\mu_i\tau_\mu=0\,,\qquad e^\mu_i e_\mu^j=\delta^j_i\,.
\end{equation}
From the $\overline{\delta}$-transformation we can deduce that the Bargmann gauge fields, including the inverse vielbeins, transform as
\begin{subequations} \label{NC_transformations_infs}
\begin{eqnarray}
\overline{\delta}\tau_{\mu} & = & \mathcal{L}_{\xi}\tau_{\mu}\\
\overline{\delta} e_{\mu}^{i} & = & \mathcal{L}_{\xi}e_{\mu}^{i}+\lambda^i{}_je_{\mu}^{j}-\lambda^{i}\tau_{\mu}\\
\overline{\delta} v^{\mu} & = & \mathcal{L}_{\xi}v^{\mu}-e_{i}^{\mu}\lambda^{i}\\
\overline{\delta} e_{i}^{\mu} & = & \mathcal{L}_{\xi}e_{i}^{\mu}+\lambda_{i}{}^{j}e_{j}^{\mu}\\
\overline{\delta} M_{\mu} & = & \mathcal{L}_{\xi}M_{\mu}+\partial_\mu \sigma-e_{\mu}^{i}\lambda_{i}\,.\label{eq:M_infinitesimal_transformation}\\
\overline{\delta}\Omega_{\mu}{}^{i} & = & \mathcal{L}_{\xi}\Omega_{\mu}{}^{i}-\partial_{\mu}\lambda^{i}+\lambda^i{}_j\Omega_{\mu}{}^{j}-\lambda^{j}\Omega_{\mu j}{}^{i}\\
\overline{\delta}\Omega_{\mu}{}^{ij} & = & \mathcal{L}_{\xi}\Omega_{\mu}{}^{ij}+\partial_{\mu}\lambda^{ij}-2\lambda^{[i}{}_k\Omega_{\mu}{}^{j]k}\,.
\end{eqnarray}
\end{subequations}

In the above $\lambda^{i}$ is the local Galilean boost parameter, $\lambda^{ij}$ the local rotation parameter and $\xi^\mu$ the generator of diffeomorphisms. The unique choice of covariant derivatives that transform covariantly under general coordinate transformations as well as under the tangent space transformations are given by:
\begin{subequations}
\begin{eqnarray}
\mathcal{D}_{\rho}\tau_{\mu} & = & \partial_{\rho}\tau_{\mu}-\Gamma_{\rho\mu}^{\lambda}\tau_{\lambda}\\
\mathcal{D}_{\rho}e_{\mu}^{i} & = & \partial_{\rho}e_{\mu}^{i}-\Gamma_{\rho\mu}^{\lambda}e_{\lambda}^{i}-\Omega_{\rho}{}^{i}\tau_{\mu}-\Omega_{\rho}{}^i{}_je_{\mu}^{j}\\
\mathcal{D}_{\rho}v^{\mu} & = & \partial_{\rho}v^{\mu}+\Gamma_{\rho\lambda}^{\mu}v^{\lambda}-\Omega_{\rho}{}^{i}e_{i}^{\mu}\\
\mathcal{D}_{\rho}e_{i}^{\mu} & = & \partial_{\rho}e_{i}^{\mu}+\Gamma_{\rho\lambda}^{\mu}e_{i}^{\lambda}+\Omega_{\rho}{}^j{}_ie_{j}^{\mu}\\
\mathcal{D}_{\rho}M_{\mu} & = & \partial_{\rho}M_{\mu}-\Gamma_{\rho\mu}^{\lambda}M_{\lambda}-\Omega_{\rho i}e_{\mu}^{i}\,.
\end{eqnarray}
\end{subequations}
The connections $\Omega_{\mu}{}^{i}$ and $\Omega_{\mu}{}^{ij}$ play the role of frame gauge fields. They are the Galilean analogue of the spin connection in general relativity. The gauge field $M_\mu$, also known as the TNC vector field, corresponds to the $U(1)$ mass gauge field in Bargmann theories. More details can be found in the references given above.

\subsection{Connections in Newton-Cartan geometry} \label{gal_connections_NC}

The vielbein postulates are
\begin{equation}
\mathcal{D}_{\rho}\tau_{\mu}=0\,,\qquad \mathcal{D}_{\rho}e_{\mu}^{i}=0\,,
\end{equation}
implying that $\mathcal{D}_{\rho}v^{\mu}=0$ and $\mathcal{D}_{\rho}e_{i}^{\mu}=0$. 
From the vielbein postulates it follows that $\tau_\mu$ and $h^{\mu\nu}=\delta^{ij}e^\mu_i e^\nu_j$ are covariantly constant, i.e.
\begin{equation}\label{eq:MC}
\nabla_\rho\tau_\mu=0\,,\qquad\nabla_\rho h^{\mu\nu}=0\,,
\end{equation}
where $\nabla_\rho$ is the covariant derivative associated with the affine connection $\Gamma_{\rho\mu}^{\lambda}$. In \cite{Hartong:2015zia} (see also  \cite{Bekaert:2014bwa})  the most general metric compatible (in the sense of \eqref{eq:MC}) affine connection was constructed
 and the result is given by
\begin{subequations}\label{HO_param}
\begin{eqnarray}
\Gamma_{\mu\nu}^{\lambda} & = & {-v^{\lambda}\partial_{\mu}\tau_{\nu}+\frac{1}{2}h^{\lambda\sigma}\left(\partial_{\mu}h_{\nu\sigma}+\partial_{\nu}h_{\mu\sigma}-\partial_{\sigma}h_{\mu\nu}\right)}+W_{\mu\nu}^{\lambda}\\
W_{\mu\nu}^{\lambda} & = & \frac{1}{2}h^{\lambda\sigma}\left(\tau_{\mu}K_{\sigma\nu}+\tau_{\nu}K_{\sigma\mu}+L_{\sigma\mu\nu}\right)\\
K_{\mu\nu} & = & -K_{\nu\mu}\,,\qquad L_{\sigma\mu\nu}=-L_{\nu\mu\sigma}\,,
\end{eqnarray}
\end{subequations}
where $K_{\mu\nu}$ and $L_{\sigma\mu\nu}$ transform as tensors under general coordinate transformations. They can be chosen arbitrarily as long as they satisfy certain transformation properties under Galilean boosts in order to leave the affine connection boost invariant. 

It follows that TNC connections \eqref{HO_param} generically have a nonzero torsion because for any $K$ and $L$ we have
\begin{equation}
2\Gamma_{[\mu\nu]}^{\lambda}\tau_\lambda=\partial_\mu\tau_\nu-\partial_\nu\tau_\mu\,.
\end{equation}
We distinguish three cases \cite{Christensen:2013lma}: Newton--Cartan (NC) geometry for which the torsion vanishes because $\tau_\mu=\partial_\mu\tau$, twistless torsional Newton--Cartan (TTNC) geometry for which $\tau_\mu=N\partial_\mu\tau$ so that $h^{\mu\rho}h^{\nu\sigma}(\partial_\mu\tau_\nu-\partial_\nu\tau_\mu)=0$ and torsional Newton--Cartan (TNC) geometry for which there are no constraints imposed on $\tau_\mu$.

It is useful to define
\begin{equation}
\Gamma_{(0)\mu\nu}^{\lambda}=-v^{\lambda}\partial_{\mu}\tau_{\nu}+\frac{1}{2}h^{\lambda\sigma}\left(\partial_{\mu}h_{\nu\sigma}+\partial_{\nu}h_{\mu\sigma}-\partial_{\sigma}h_{\mu\nu}\right)\,,\label{eq:pseudo_connection}
\end{equation}
so that we can write the affine connection as $\Gamma_{\mu\nu}^{\lambda}=\Gamma_{(0)\mu\nu}^{\lambda}+W_{\mu\nu}^{\lambda}$.
The object $\Gamma_{(0)\mu\nu}^{\lambda}$ is expressed in terms of vielbeins or metric quantities only and transforms correctly, i.e. as an affine connection under GCTs, but it is not invariant under local Galilean boosts. In order to construct a Galilean boost invariant connection we must take $W_{\mu\nu}^{\lambda}$ to be non-zero. In analogy to Riemann--Cartan geometry where we know that any connection can be written as the Levi-Civita connection plus the contortion tensor it is useful to think of $\Gamma_{(0)\mu\nu}^{\lambda}$ as a (specific) \textit{pseudo}-connection and $W_{\mu\nu}^{\lambda}$ as a \textit{pseudo}-contortion tensor.

Likewise from the vielbein postulates it follows that we can write the connections $\Omega_{\mu i}$ and $\Omega_{\mu ij}$  as the sum of two terms
\begin{subequations}
\label{eq:connectionsaa}
\begin{eqnarray}
\Omega_{\mu i} & = & \Omega_{(0)\mu i}+C_{\mu i}\\
\Omega_{\mu ij} & = & \Omega_{(0)\mu ij}+C_{\mu ij}\,,
\end{eqnarray}
\end{subequations}
where
\begin{subequations}\label{eq:Pseudo_Levicivitalike_connection}
\begin{eqnarray}
\Omega_{(0)\mu i}  & = & v^{\nu}\partial_{[\nu}e_{\mu]}^{i}+v^{\nu}e^{\sigma i}e_{\mu j}\partial_{[\nu}e_{\sigma]}^{j}\\
\Omega_{(0)\mu ij} & = & e_{[i|}^{\lambda}\partial_{\lambda}e_{\mu|j]}-e_{[i|}^{\lambda}\partial_{\mu}e_{\lambda|j]}-e_{\mu k}e_{[i}^{\sigma}e_{j]}^{\lambda}\partial_{\lambda}e_{\sigma}^{k}\,,\label{eq:rotation_gauge_field_special_Gal}\\
\nonumber\\
C_{\mu i} & = & -v^{\nu}e_{\lambda i}W_{\mu\nu}^{\lambda}\\
C_{\mu ij} & = & e_{j}^{\nu}e_{\lambda i}W_{\mu\nu}^{\lambda}\,.
\end{eqnarray}
\end{subequations}

Any choice of pseudo-contortion tensor $W_{\mu\nu}^{\lambda}$ with the right transformation properties leads to a good TNC connection.
In particular there exists a unique connection linear in the mass gauge field $M_{\mu}$ known in the TNC literature \cite{Jensen:2014aia,Hartong:2014pma,Bergshoeff:2014uea}. In the above parameterization this is given by
\begin{subequations}
\label{eq:minconnho}
\begin{eqnarray}
K_{\sigma\rho} & = & 2\partial_{[\sigma}M_{\rho]}\\
L_{\sigma\mu\nu} & = & 2M_{\sigma}\partial_{[\mu}\tau_{\nu]}-2M_{\mu}\partial_{[\nu}\tau_{\sigma]}+2M_{\nu}\partial_{[\sigma}\tau_{\mu]}\,.
\end{eqnarray}
\end{subequations}
Except for the case of vanishing torsion the tensor $L_{\sigma\mu\nu}$ is not invariant under the mass $U(1)$ gauge transformation with parameter $\sigma$. (It is not possible to construct out of the vielbeins and $M_\mu$ a metric compatible connection \eqref{eq:MC} that is invariant under Galilean boosts and $U(1)$ mass gauge transformations \cite{Jensen:2014aia,Hartong:2014pma,Bergshoeff:2014uea}.) The affine connection associated with these choices of $K_{\sigma\rho}$ and $L_{\sigma\mu\nu}$ is denoted by $\bar\Gamma^\lambda_{\mu\nu}$ and reads
\begin{subequations}
\begin{eqnarray}
\bar\Gamma_{\mu\nu}^{\lambda} & = & -\hat{v}^{\lambda}\partial_{\mu}\tau_{\nu}+\frac{1}{2}h^{\lambda\sigma}\left(\partial_{\mu}\overline{h}_{\nu\sigma}+\partial_{\nu}\overline{h}_{\mu\sigma}-\partial_{\sigma}\overline{h}_{\mu\nu}\right) \label{gravphotonic_conn}\\
\hat{v}^{\mu} & = &	v^{\mu}-h^{\mu\lambda}M_{\lambda}\\ 
\overline{h}_{\mu\nu} & = & h_{\mu\nu}-2\tau_{(\mu}M_{\nu)}\,.
\end{eqnarray}
\end{subequations}
We shall prove using the Noether procedure that $\bar\Gamma_{\mu\nu}^{\lambda}$ is, in a sense to be made more precise below, the minimal TNC connection.


\section{The Noether procedure for Galilean theories} \label{noether_gal}

We start applying the Noether procedure to theories with Galilean symmetries without the massive Bargmann extension.
We do not need to assume that the theories have a Lagrangian description.
In section \ref{conserved_noether_gal} we analyze the symmetry currents for all the generators of the Galilei algebra.
It will be convenient to consider improvements of these currents that simplify the structure of the Galilean current multiplet.
For example the stress tensor can be made symmetric using an improvement transformation. This will be the subject of section \ref{gal current improvements}.
The actual Noether procedure is then discussed in section \ref{Galilean_noether}. Here we introduce all the gauge connections needed to make the theory invariant under local Galilean transformations at lowest order. The gauge fields will couple to the improved currents and this has some interesting consequences for the construction of dependent connections. Finally in section \ref{The minimal Galilean connection} we discuss how the Noether procedure gives rise to an affine (linearized) Christoffel-type minimal connection.

\subsection{Conserved symmetry currents} \label{conserved_noether_gal}

Consider a local field theory that is invariant under the Galilean transformations \eqref{eq:Galilean_group}.
There exist conserved currents $E^\mu,\,T^{\mu i},\,b^{\mu i},\,j^{\mu ij}$ whose $\mu=0$ components integrate to time-independent charges satisfying the algebra \eqref{eq:Galilean_algebra} 
\begin{subequations}
\begin{eqnarray}
H & = & \int\mathrm{d}^d x \,E^0\\
P^i & = & \int\mathrm{d}^d x \,T^{0i}\\
G^i & = & \int\mathrm{d}^d x \,b^{0i}\\
J^{ij} & = & \int\mathrm{d}^d x \,j^{0ij}\,.
\end{eqnarray}
\end{subequations}
$E^\mu,\,T^{\mu i},\,b^{\mu i},\,j^{\mu ij}$ are the energy, momentum, boost and rotation currents respectively and constitute the general current multiplet of any Galilean invariant theory perhaps along with currents for additional symmetries. In particular $T^{ij}$ is the spatial stress tensor, which at this stage is not necessarily symmetric.

The commutation relations of  $B^i,\,J^{ij}$ with the translation generators $H,\,P^i$ imply that $b^{\mu i},\,j^{\mu ij}$ are of the form
\begin{subequations}\label{Galilei_conserved_currents_general}
\begin{eqnarray}
b^{\mu i} & = & t\,T^{\mu i}+w^{\mu i}\\
j^{\mu ij} & = & x^{i}T^{\mu j}-x^{j}T^{\mu i}+s^{\mu ij}\,,
\end{eqnarray}
\end{subequations}
where $w^{\mu i}$ and $s^{\mu ij}=-s^{\mu ji}$ are local and do not explicitly depend on the coordinates but generically are not conserved. 
We call the non-conserved current $w^{\mu i}$ the \emph{internal boost-current}
while we refer to $s^{\mu ij}=-s^{\mu ji}$ as the \emph{spin-current}.

The conservation laws for the boost and rotation currents in the form \eqref{Galilei_conserved_currents_general} lead, on shell, to the following identities involving $T^{\mu i}$:
\begin{subequations}\label{eq:galilean_no_masscurrent}
\begin{eqnarray}
\partial_{\mu}b^{\mu i}=0\qquad & \Rightarrow &\qquad T^{0i}  =  -\partial_{\mu}w^{\mu i} \label{gal_mon_dens}\\
\partial_{\mu}j^{\mu ij}=0\qquad & \Rightarrow &\qquad 2T^{[ij]}  =  -\partial_{\mu}s^{\mu ij}\label{eq:antisym_T}\,.
\end{eqnarray}
\end{subequations}
This shows that the stress tensor $T^{ij}$ is only
symmetric if $s^{\mu ij}$ is conserved or trivial.

\subsubsection*{Canonical Noether currents}

We can give explicit expressions for the currents when provided with a Lagrangian density $\mathcal{L}\left(\varphi_{\ell},\partial\varphi_{\ell}\right)$.
Here we denote the field content of the theory by $\varphi_{\ell}$
where the index $\ell$ distinguishes various
internal components. In particular we assume that the $\varphi_{\ell}$ transform linearly under
the homogeneous part $\mathbb{R}^{d}\ltimes\mathrm{SO}\left(d\right)$ of the Galilean group.
\begin{subequations}\label{eq: Galilean transform}
\begin{eqnarray}
&& x^{\prime\mu}  \quad\;\, =  \;x^{\mu}+\delta x^{\mu},\\
&& \delta t  \quad \;\;\;=\;  \epsilon_{0},\\
&& \delta x_{i}  \quad \;\, =  \;\epsilon_{i}+\lambda_{i}t+\lambda_{ij}x^{j}\,,\\
&& \varphi_{\ell}^{\prime}\left(x^{\prime}\right)  =  \;\varphi_{\ell}\left(x\right)+\delta\varphi_{\ell}\left(x\right),\\
&& \delta\varphi_{\ell}\left(x\right) \! =  \;\lambda^{i}\left(G_{i}\right)_{\ell\ell^{\prime}}\varphi_{\ell^{\prime}}\left(x\right)+\frac{1}{2}\lambda^{ij}\left(J_{ij}\right)_{\ell\ell^{\prime}}\varphi_{\ell^{\prime}}\left(x\right)\,, \label{eq:varphi_gal}
\end{eqnarray}
\end{subequations}
where $\lambda^{ij}=-\lambda^{ji}$ are infinitesimal spatial
rotation parameters, $\lambda^{i}$ boost parameters and $\epsilon^{0},\,\epsilon^{i}$
infinitesimal translation parameters that together constitute all the Galilei transformations. The generators of rotations $\left(J_{ij}\right)_{\ell\ell^{\prime}}$
are in a suitable linear representation of $\mathrm{SO}\left(d\right)$ realized
on the fields $\varphi_{\ell}$, and $\left(G_{i}\right)_{\ell\ell^{\prime}}$
furnishes a linear representation specifying how the Galilean boosts act on the fields.
One can vary the corresponding Galilean-invariant action $S^{\left(0\right)}\left[\varphi\right]=\int_{M}\mathrm{d}^{D}x\mathcal{L}\left(\varphi_{\ell},\partial\varphi_{\ell}\right)$
directly to derive the canonical Noether currents \cite{Kibble:1961ba}
\begin{subequations}
\label{Galilei_conserved_currents}
\begin{eqnarray}
&&E_{\mathrm{can}}^{\mu}  =  \frac{\partial\mathcal{L}}{\partial\left[\partial_{\mu}\varphi_{\ell}\right]}\partial_{0}\varphi_{\ell}-\delta_{0}^{\mu}\mathcal{L}\\
&&T_{\mathrm{can}}^{\mu i}  =  \frac{\partial\mathcal{L}}{\partial\left[\partial_{\mu}\varphi_{\ell}\right]}\partial^{i}\varphi_{\ell}-\delta^{\mu i}\mathcal{L}\\
&&b_{\mathrm{can}}^{\mu i}  =  tT_{\mathrm{can}}^{\mu i}+w^{\mu i}\\
&&j_{\mathrm{can}}^{\mu ij}  =  x^{i}T_{\mathrm{can}}^{\mu j}-x^{j}T_{\mathrm{can}}^{\mu i}+s^{\mu ij}\,,
\end{eqnarray}
\end{subequations}
where we defined
\begin{subequations}
\begin{eqnarray}
&&w^{\mu i} = -\frac{\partial\mathcal{L}}{\partial\left[\partial_{\mu}\varphi_{\ell}\right]}\left(G^{i}\right)_{\ell\ell^{\prime}}\varphi_{\ell^{\prime}}\\
&&s^{\mu ij} = -\frac{\partial\mathcal{L}}{\partial\left[\partial_{\mu}\varphi_{\ell}\right]}\left(J^{ij}\right)_{\ell\ell^{\prime}}\varphi_{\ell^{\prime}}\,.
\end{eqnarray}
\end{subequations}
The currents \eqref{Galilei_conserved_currents} are conserved on shell as required.

\subsection{Improvement of the currents} \label{gal current improvements}

The symmetry currents $E^\mu,\,T^{\mu i},\,b^{\mu i},\,j^{\mu ij}$ of a general Galilean theory are only defined up to a total divergence
that can be added without changing the conserved charges.
Hence, one can always improve the current multiplet of the theory
as follows
\begin{subequations}\label{eq:Improved currents}
\begin{eqnarray}
&&E_{\mathrm{imp}}^{\mu}  =  E^{\mu}+\partial_{\rho}A^{\rho\mu}\\
&&T_{\mathrm{imp}}^{\mu i} \, =  T^{\mu i}+\partial_{\rho}B^{\rho\mu i} \label{Timp0} \\
&&b_{\mathrm{imp}}^{\mu i}  \,\,=  b^{\mu i}+\partial_{\rho}E^{\rho\mu i}\\
&&j_{\mathrm{imp}}^{\mu ij} \,\, =  j^{\mu ij}+\partial_{\rho}D^{\rho\mu ij}\,,
\end{eqnarray}
\end{subequations}
where all the improvement terms are antisymmetric in $\mu,\rho$.
The choice of improvements, can be used to make the currents simpler.
For instance in the case of relativistic theories, the Belinfante-Rosenfeld procedure (which we review in appendix \ref{currents_lorentz})
allows to define a symmetric and gauge invariant energy-momentum tensor (see for instance \cite{Dumitrescu:2011iu}).

In our case we want to do something similar and construct improvements such that the currents corresponding to Galilean boosts and rotations are as simple as possible. This is achieved by parameterizing the improvements in \eqref{eq:Improved currents} in terms of $s^{\mu ij}$ and $w^{\mu i}$, in an essentially unique way\footnote{There are further improvements of the momentum current that leave $T_{\mathrm{imp}}^{[ij]}=0$. These can be performed by adding an improvement term $\partial_{\rho}\tilde{B}^{\rho\mu i}$ with $\tilde{B}^{\rho j i}=\tilde{B}^{\rho i j}$ to the momentum current.} to simplify the conservation equations $\partial_{\mu}j_{\mathrm{imp}}^{\mu ij}=\partial_{\mu}b_{\mathrm{imp}}^{\mu i}=0$ the most:
\begin{subequations}\label{eq:improvements_galilean}
\begin{eqnarray}
&&B^{\rho \mu i}  \,\,=  2\delta_{k}^{[\mu}\delta_{0}^{\rho]}\left(w^{(ik)}+\frac{1}{2}s^{0ik}\right)-\frac{1}{2}\delta_{j}^{\mu}\delta_{k}^{\rho}\left(s^{kji}+s^{ijk}+s^{jik}\right) \label{Timp1} \\
&&D^{\rho \mu ij}  =  x^{i}B^{\rho \mu j}-x^{j}B^{\rho \mu i}\\
&&E^{\rho \mu i}  \,\, =  tB^{\rho \mu i}\,.
\end{eqnarray}
\end{subequations}
This leads to
\begin{subequations}\label{Galilei_improved_currents_general}
\begin{eqnarray}
b_{\mathrm{imp}}^{\mu i} & = & tT_{\mathrm{imp}}^{\mu i}+\psi^{\mu i}\\
j_{\mathrm{imp}}^{\mu ij} & = & x^{i}T_{\mathrm{imp}}^{\mu j}-x^{j}T_{\mathrm{imp}}^{\mu i}\,,
\end{eqnarray}
\end{subequations}
where we have defined the current
\begin{equation}
\psi^{\mu i}=\delta_{0}^{\mu}w^{0i}+{1\over 2}\delta_{j}^{\mu}\left(w^{ji}-w^{i j}-s^{0ij}\right)\,.\label{eq:psi_current}
\end{equation}
The conservation laws for the currents \eqref{Galilei_improved_currents_general} now give the following:
\begin{subequations}
\begin{eqnarray}
T_{\mathrm{imp}}^{0i} & = & -\partial_{\mu}\psi^{\mu i} \label{impr_gal_mon_dens}\\
T_{\mathrm{imp}}^{[ij]} & = & 0\,.
\end{eqnarray}
\end{subequations}

Hence the stress tensor can always be made symmetric, and $T_{\mathrm{imp}}^{0i}$ is the total derivative of some generically non-conserved current $\psi^{\mu i}$ that is antisymmetric in its spatial indices. It is important to note that the current $\psi^{\mu i}$ is the only combination of $w^{\mu i}$ and $s^{\mu ij}$ that remains in any of the symmetry currents.

\subsection{Coupling the gauge fields to currents} \label{Galilean_noether}

Consider a theory in flat space that is invariant under global Galilean transformations. Let it be described by the action functional $S^{\left(0\right)}\left[\varphi\right]$. In this section we proceed to gauge the Galilean group at linearized level. 
The variation of $S^{\left(0\right)}\left[\varphi\right]$ under local translations, boosts and rotations reads:
\begin{equation}
\delta S^{\left(0\right)}\left[\varphi\right]=-\int_{M}\mathrm{d}^{D}x\left(\partial_{\mu}\epsilon_{0}E^{\mu}+\partial_{\mu}\epsilon_{i}\,T^{\mu i}+\partial_{\mu}\lambda_{i}b^{\mu i}+\frac{1}{2}\partial_{\mu}\lambda_{ij}j^{\mu ij}\right)\,.\label{eq:varivari}\end{equation}
The expression above vanishes when the parameters do not depend on position as required from invariance under global Galilean transformations.  It vanishes on-shell for any choice of the parameters by virtue of current conservation. It is useful to introduce the infinitesimal parameters
$\xi_{0} = \epsilon_{0} $,  
 $\xi_{i}=\epsilon_i +t \lambda_i-\lambda_{i j} x^j$, which we will relate to temporal/spatial diffeomorphisms  in the following, and rewrite the variation of the action as
\begin{equation}
\delta S^{\left(0\right)}\!\left[\varphi\right]=\!-\!\int_{M}\!\!\mathrm{d}^{D}\!x\left(\partial_{\mu}\xi_{0}E^{\mu}\!+(\partial_{\mu}\xi_{i}-\!\lambda_i \delta^0_\mu+\!\lambda_{i j} \delta^j_\mu)\,T^{\mu i}\!+\partial_{\mu}\lambda_{i}w^{\mu i}+\frac{1}{2}\partial_{\mu}\lambda_{ij}s^{\mu ij}\right).\label{eq:Variation_local_galilean}
\end{equation}

Next we introduce gauge fields  $\overline{\tau}_{\mu},\,\overline{e}_{\mu i}$ and  $\overline{\Omega}_{\mu i},\,\overline{\Omega}_{\mu ij}$ that transform as follows under local Galilean transformations:
\begin{subequations} \label{eq:transformations_linear_fields}
\begin{eqnarray}
&&\delta^{\left(1\right)}\overline{\tau}_{\mu}  \,\,\,=  \,\partial_{\mu}\xi_{0}\\
&&\delta^{\left(1\right)}\overline{e}_{\mu i}  \,\,=  \,\partial_{\mu}\xi_{i}-\lambda_{i}\delta_{\mu}^{0}+\lambda_{ij}\delta_{\mu}^{j}\label{eq:inttrs}\\
&&\delta^{\left(1\right)}\overline{\Omega}_{\mu i} \, = \, -\partial_{\mu}\lambda_{i}\\
&&\delta^{\left(1\right)}\overline{\Omega}_{\mu ij}  \!=\,  \partial_{\mu}\lambda_{ij}\,.
\end{eqnarray}
\end{subequations}
Here the notation $\delta^{(1)}$ indicates that these expressions are valid at first order in the variation parameters and the gauge field and will be modified at higher orders.

We can add to the action $S^{\left(0\right)}\left[\varphi\right]$ couplings of the gauge fields to the components of the current multiplet as follows:
\begin{equation}
S\left[\varphi;\overline{e},\overline{\tau},\overline{\Omega}_i,\overline{\Omega}_{ij}\right]= S^{\left(0\right)}\left[\varphi\right]+S^{\left(1\right)}\left[\varphi;\overline{e},\overline{\tau},\overline{\Omega}_i,\overline{\Omega}_{ij}\right]\,, \label{action_noether}
\end{equation}
where
\begin{equation}
S^{\left(1\right)}=\int_{M}\mathrm{d}^{D}x\,\left[\overline{\tau}_{\mu}E^{\mu}+\overline{e}_{\mu i}T^{\mu i}-\overline{\Omega}_{\mu i}w^{\mu i}+\frac{1}{2}\overline{\Omega}_{\mu ij}s^{\mu ij}\right]\,.\label{eq:gauge_couple_galilean}
\end{equation}
Under a local Galilean transformation the variation of $S^{\left(0\right)}$ given by \eqref{eq:Variation_local_galilean} is cancelled by terms coming from the variation of the gauge fields in $S^{\left(1\right)}$ even without use of the equations of motion. The modified action is therefore invariant off shell at first order. On shell the variation of $S^{\left(0\right)}\left[\varphi\right]$ is zero and invariance of \eqref{eq:gauge_couple_galilean} follows from current conservation ($\partial_\mu T^{\mu i}=0,\;\partial_\mu E^\mu=0$ and  \eqref{eq:galilean_no_masscurrent}). The coupling of some gauge fields to non-conserved currents in \eqref{eq:gauge_couple_galilean} is a special feature
of gauging spacetime symmetries and can be traced back to the non-derivative terms in the variation of $\bar e_{\mu i}$ as in  \eqref{eq:inttrs}.\footnote{When interpreting $\bar e_{\mu i}$ as a linearized (spatial) vielbein these terms can be traced back to the fact that in flat space (at zero$^{\rm{th}}$ order) the vielbein is not invariant under Galilean frame rotations (see appendix \ref{sec:lin_NC} for details).}

In principle one could go to higher orders in the gauge fields and determine the transformation laws and terms to add to the action to have local Galilean invariance at all orders. However, beyond the linear level this procedure is not unique and rather cumbersome.\footnote{ See \cite{Kraus:1992gk} for a discussion of these issues in a relativistic framework.} We will confine ourselves to the  analysis at linear order.

In section \ref{sec:Comparison to usual procedure} we will consider the precise relation between the results of the Noether procedure and torsional Newton-Cartan geometry. At this stage the gauge fields are a priori unrelated to the TNC objects that we presented in section \ref{TNC_gauge}. However we will see that their relation is exactly as implied by the notation we used.

Next we consider how improvements of the current multiplet affect the discussion above. In the variation of $S^{(0)}$ given by \eqref{eq:varivari} improvements are inconsequential as they give rise to boundary terms that vanish in flat space. We can then cancel the variation of $S^{(0)}$ coupling the gauge fields to the currents with any choice of improvement. In particular we can select \eqref{eq:improvements_galilean} which leads to:
\begin{equation}
S^{\left(1\right)}=\int_{M}\mathrm{d}^{D}x\,\left[\overline{\tau}_{\mu}E^{\mu}+\overline{e}_{\mu 0}T_{\rm{imp}}^{0 i}+\overline{s}_{i j}T_{\rm{imp}}^{i j}-\overline{\Omega}_{\mu i}\psi^{\mu i}\right]\,,\label{eq:gauge_couple_impr}
\end{equation}
where $\overline{s}_{ij}=\overline{e}_{(i j)}$ plays the role of a linearized spatial metric coupling to the symmetric energy momentum tensor.%
\footnote{Note that the improved (spatial) stress tensor $T_{\rm{imp}}^{i j}$  defined via \eqref{Timp0}, \eqref{Timp1} is symmetric on shell.}
The discussion above implies that the difference between \eqref{eq:gauge_couple_impr} and \eqref{eq:gauge_couple_galilean} is invariant under local Galilean transformation at first order. In the next section we will explore the consequences of this fact. Before proceeding however, we notice that in order to cancel the variation of $S^{(0)}$ coupling $\psi^{\mu i}$ to $\overline{\Omega}_{\mu i}$ in \eqref{eq:gauge_couple_impr} is overkill due to the fact that $\psi^{i j}=-\psi^{ j i}$. Indeed we have:
\begin{equation}
\label{eq:couplpsi}
\int_{M}\mathrm{d}^{D}x\,\overline{\Omega}_{\mu i}\psi^{\mu i}=\int_{M}\mathrm{d}^{D}x\,\left[\overline{\Omega}_{0 i}\psi^{0 i}+{1\over 2}( \overline{\Omega}_{i j}- \overline{\Omega}_{j i})\psi^{i j}\right]~,
\end{equation}
and the variations of the relevant combinations of $\overline{\Omega}$ are
\begin{equation}
\delta^{(1)}( \overline{\Omega}_{i j}- \overline{\Omega}_{j i})=\partial_j \lambda_i-\partial_i \lambda_j~,\qquad  \delta^{(1)} \overline{\Omega}_{0 i}= - \partial_0 \lambda_i~. 
\end{equation}
Hence we can obtain the same result replacing $\overline{\Omega}_{\mu i}$ with the curl of a one form $\overline{ M}_\mu$:
\begin{equation}
\label{eq:couplpsitwo}
\int_{M}\mathrm{d}^{D}x\,(\partial_\mu \overline{ M}_i -\partial_i \overline{ M}_\mu)\psi^{\mu i}~,
\end{equation}
where the first order variation of $\overline{ M}_\mu$ under a Galilean transformation is 
\begin{equation}
\delta^{\left(1\right)}\overline{M}_{\mu}=-\delta_{\mu}^{i}\lambda_{i}\,.
\end{equation}
The coupling of $\overline{ M}_\mu$ is also invariant under local $U(1)$ transformations that only act on $\overline{ M}_\mu$ as $\delta \overline{ M}_\mu=-\partial_\mu \sigma$. Because this extra $U(1)$ does not act on the fields in $S^{(0)}$ its presence does not constrain the theory under consideration.
Hence we can replace $S^{\left(1\right)}$ in \eqref{eq:gauge_couple_impr} with
\begin{eqnarray}
S^{\left(1\right)}=&&\int_{M}\mathrm{d}^{D}x\,\left[\overline{\tau}_{\mu}E^{\mu}+\overline{e}_{0 i}T_{\rm{imp}}^{0 i}+\overline{s}_{ i j}T_{\rm{imp}}^{ i j}-(\partial_\mu \overline{ M}_i -\partial_i \overline{ M}_\mu)\psi^{\mu i}\right]\cr=
&&\int_{M}\mathrm{d}^{D}x\,\left[\overline{\tau}_{\mu}E^{\mu}+\overline{e}_{0 i}T_{\rm{imp}}^{0 i}+\overline{s}_{ i j}T_{\rm{imp}}^{ i j}+\overline{ M}_\mu\Phi^{\mu}\right]\,,\label{eq:gauge_couple_imprtwo}
\end{eqnarray}
where in the last line we integrated by parts and introduced:
\begin{equation}
\Phi^{\rho}=-\left(\partial_{0}\psi^{0j}+\partial_{i}\psi^{ij}\right)\delta_{j}^{\rho}+\partial_{j}\psi^{0j}\delta_{0}^{\rho}\,.\label{eq:topological_current_noether}
\end{equation}
The current $\Phi^{\rho}$ is identically conserved by the antisymmetry of $\psi^{ij}$. Its conserved charge is zero in flat space using Stokes' theorem and in this sense $\Phi^{\rho}$ is a \emph{topological current}.

We argue that $S^{\left(1\right)}$ in the form \eqref{eq:gauge_couple_imprtwo} is the minimal coupling to gauge fields that is generically necessary to ensure invariance of $S^{\left(0\right)}+S^{\left(1\right)}$ under local Galilean transformations at linear order.

\subsection{The minimal affine connection} \label{The minimal Galilean connection}

Next we evaluate the difference between \eqref{eq:gauge_couple_imprtwo} and \eqref{eq:gauge_couple_galilean}. Integrating by parts we can recast it as\begin{equation}
\int_{M}\mathrm{d}^{D}x\,\left[\overline{Y}_{\mu i}w^{\mu i}-\frac{1}{2}\overline{Y}_{\mu ij}s^{\mu ij}\right]\,, \label{eq:action_contortions_Gal}
\end{equation}
The two objects $\overline{Y}_{\mu i}$ and $\overline{Y}_{\mu ij}$ are invariant under local Galilean transformations at first order. Hence they are the linearization about flat space of tensors on $M$.  Their explicit expressions are given by:
\begin{subequations} \label{noether_pseudo_conn}
\begin{equation}
\overline{Y}_{\mu ij}=\overline{\Omega}_{\mu ij}-\tilde{\Omega}_{ \mu ij}\,,\qquad\overline{Y}_{\mu i}=\overline{\Omega}_{\mu i}-\tilde{\Omega}_{\mu i}\,,
\end{equation}
\begin{eqnarray}
&&\tilde{\Omega}_{\mu i}\;=-\delta_{\mu}^{k}\left(\frac{1}{2}\partial_{0}\overline{s}_{ki}+\partial_{(k}\overline{v}_{i)}\right)-2\delta_{\mu}^{0}\partial_{[0}\overline{M}_{i]}-\delta_{\mu}^{j}\partial_{[j}\overline{M}_{i]}\,,\\
&&\tilde{\Omega}_{ \mu ij}=-\delta_{\mu}^{0}\left(\partial_{0}\overline{e}_{[ij]}+\partial_{[i}\overline{v}_{j]}-\partial_{[i}\overline{M}_{j]}\right)+\delta_{\mu}^{k}\left(-\partial_{k}\overline{e}_{[ij]}+\partial_{[i}\overline{s}_{j]k}\right)\,,
\end{eqnarray}
\end{subequations}
where we defined $\bar v_i=-\bar e_{0i}$ while $\bar s_{i j}=\overline{e}_{(i j)}$ as above.
It can be checked explicitly that $\tilde{\Omega}_{\mu i}$ and $\tilde{\Omega}_{\mu ij}$ transform as $\overline{\Omega}_{\mu i}$ and  $\overline{\Omega}_{\mu i j}$ respectively under (first order) local Galilean transformations. Hence their interpretation is clear: they are linearized expressions for the connections $\overline{\Omega}_{\mu i}$ and  $\overline{\Omega}_{\mu i j}$ written in terms of the remaining gauge fields $\overline{\tau}_\mu,\bar e_{\mu i}$ and $\overline{M}_\mu$. 
In this respect they are equivalent to the Levi-Civita expression for the spin connection in relativistic theories. Indeed in a relativistic framework an argument parallel to the above results in the linearized Levi-Civita spin connection (see appendix \ref{sec:The-Noether-procedure-lorentz} for details). The interpretation of $\overline{Y}_{\mu i}$ and $\overline{Y}_{\mu ij}$ is then that of contortion tensors describing the coupling to the non-conserved currents $w^{\mu i}$ and $s^{\mu i j}$. 

We argued that the minimal current multiplet for generic Galilean field theories is obtained with the choice of improvements \eqref{eq:improvements_galilean}. Hence \eqref{eq:gauge_couple_imprtwo} (equivalently $\overline{Y}_{\mu i}=0$ and $\overline{Y}_{\mu ij}=0$) is the minimal gauging of Galilean symmetries. The argument above therefore singles out  $\tilde{\Omega}_{ \mu i}$ and $\tilde{\Omega}_{ \mu ij}$ as the minimal choice of connections (at first order). We can obtain any other Galilean connection by allowing nonzero "contortion"-like tensors ${\overline{Y}}_{\mu i},\,{\overline{Y}}_{\mu ij}$. This will result in connections that couple non-minimally to the internal boost and spin currents $w^{\mu i}\,,s^{\mu ij}$. We will compare these results, obtained at the linear order, with the results of Torsional Newton Cartan geometry at full nonlinear level in section \ref{sec:Comparison to usual procedure}.

\section{The Noether procedure for Bargmann theories}\label{noether_barg}

We will now consider theories with Bargmann or massive Galilean symmetries. The prototypical example of such a theory is the Schr\"odinger model for a complex scalar field whose equation of motion is the Schr\"odinger equation. The massive extension is realized by a $U(1)$ phase transformation of the complex scalar field.
In the first two sections \ref{conserved_noether_barg} and \ref{sub:Improvement_currents_Bargmann} we repeat the analysis of the previous section, but now for Bargmann theories. 

The Bargmann central extension of the Galilei algebra leads to an extra gauge connection $M_\mu$ for mass conservation. In section \ref{Bargmann_noether} we study the role of this gauge field in detail from the point of view of the Noether procedure and we give an interpretation for why, in the case of theories with massless Galilean symmetries like Galilean electrodynamics, we still need to introduce the Newton--Cartan vector $M_\mu$ to describe the coupling of the theory to a curved background. 

\subsection{Conserved symmetry currents} \label{conserved_noether_barg}
Consider first a local field theory invariant under the Bargmann group \eqref{eq:Bargmann_group}.
In addition to the currents we have in the Galilean case, there is also a conserved current $J^\mu$ corresponding to the central charge $N$.
The non-zero commutator \eqref{eq:Bargmann_algebra} implies that the form of the boost current $b^{\mu i}$ is now
\begin{equation}
b^{\mu i} = tT^{\mu i}-x^{i}J^{\mu}+w^{\mu i}\label{eq:b_current_Bargmann}\,,
\end{equation}
while the rest of the current multiplet are as in section \ref{conserved_noether_gal}.

The conservation law for $j^{\mu ij}$ still imply \eqref{eq:antisym_T}, but the new conservation law for $b^{\mu i}$ given by \eqref{eq:b_current_Bargmann} now changes significantly from \eqref{gal_mon_dens}. It now implies a relation between the flux $J^i$ and momentum density given by
\begin{equation} \label{eq:Bargmann_masscurrent}
T^{0i}  =  J^{i}-\partial_{\mu}w^{\mu i}\,.
\end{equation}

\subsubsection*{Lagrangian formulations and canonical currents}
Consider now field theory with a Lagrangian description. Bargmann representations are projective Galilean representations with the fields transforming with the projective factor \cite{Bargmann:1954gh}

\begin{equation}
\exp\left(-if\left(t,\boldsymbol{x}\right)N\right)\,,\qquad f\left(t,\boldsymbol{x}\right)=\frac{1}{2}\boldsymbol{v}^{2}t+\boldsymbol{v}^{t}\mathrm{\boldsymbol{\mathrm{R}}}\boldsymbol{x}-\sigma\,.
\end{equation}

The extra parameter compared to the Galilean case
is $\sigma\in\mathbb{R}$ corresponding to the global $U(1)$
symmetry of the field theory, while $\boldsymbol{v}$ parametrizes a finite boost and $\boldsymbol{\mathrm{R}}$ a finite rotation.
Thus the only difference to the Galilean transformations \eqref{eq: Galilean transform} of the previous section
is that now there are two extra terms in the infinitesimal variation $\delta\varphi_{\ell}$ due
to the representation of the central charge $N$ on the field components.
The infinitesimal version of the field transformations \eqref{eq:varphi_gal} is now replaced with

\begin{subequations}\label{eq:Bargmann_transform_infinitesimal}

\begin{eqnarray}
\delta\varphi_{\ell}\left(x\right) & = & +i\sigma\left(N\right)_{\ell\overline{\ell}}\varphi_{\overline{\ell}}\left(x\right)-i\lambda^{i}x_{i}\left(N\right)_{\ell\overline{\ell}}\varphi_{\overline{\ell}}\left(x\right)\nonumber \\
 &  & +\lambda^{i}\left(G_{i}\right)_{\ell\overline{\ell}}\varphi_{\overline{\ell}}\left(x\right)+\frac{1}{2}\lambda^{ij}\left(J_{ij}\right)_{\ell\overline{\ell}}\varphi_{\ell^{\prime}}\left(x\right)\,.
\end{eqnarray}

\end{subequations}

We  will refer to the conserved $U(1)$ current $J^{\mu}$ as the mass current. The
canonical currents  take the expressions
\begin{subequations}\label{Galilei_conserved_currents-1}
\begin{eqnarray}
J_{\mathrm{can}}^{\mu} & = & -i\frac{\partial\mathcal{L}}{\partial\left[\partial_{\mu}\varphi_{\ell}\right]}\left(N\right)_{\ell\overline{\ell}}\varphi_{\overline{\ell}}\\
b_{\mathrm{can}}^{\mu i} & = & tT_{\mathrm{can}}^{\mu i}-x^{i}J_{\mathrm{can}}^{\mu}+w^{\mu i}\,,
\end{eqnarray}
\end{subequations}
while the other currents remain as in \eqref{Galilei_conserved_currents}. 

It is useful to write the variation of the action under local Bargmann transformations
\begin{equation}
\delta S\left[\varphi\right]=-\int_{M} [\,\partial_{\mu}\xi_{0}E_{\mathrm{can}}^{\mu}+(\partial_{\mu}\xi_{i}-\!\lambda_i \delta^0_\mu+\!\lambda_{i j} \delta^j_\mu)T_{\mathrm{can}}^{\mu i}+\partial_{\mu}\sigma J_{\mathrm{can}}^{\mu}+\partial_{\mu}\lambda_{i}w^{\mu i}+\frac{1}{2}\partial_{\mu}\lambda_{ij}s^{\mu ij}]\,,\label{eq:Variation_local_bargmann}
\end{equation}
As in the Galilean case, modifying the currents above by improvement terms would be of no consequence in flat space.

\subsection{Improvement of the currents}\label{sub:Improvement_currents_Bargmann}

The new symmetry current $J^\mu$ brings in a new improvement transformation of the current multiplet besides those of \eqref{eq:Improved currents} 
\begin{equation}
J_{\mathrm{imp}}^{\mu}=J^{\mu}+\partial_{\rho}C^{\rho\mu}\,.
\end{equation}
We can now choose the new improvement $C^{\rho\mu}=-C^{\mu\rho}$ of $J^\mu$ to simplify the constraint on the momentum current \eqref{eq:Bargmann_masscurrent}
as much as possible. The improvement $E^{\rho\mu i}$ of the boost current $b^{\mu i}$ must be chosen in a different way than in the Galilean case \eqref{eq:improvements_galilean}, but the remaining improvements are identical. The currents are maximally simplified by
\begin{eqnarray}
C^{\rho\mu} & = & 2\delta_{i}^{[\rho}\delta_{0}^{\mu]}w^{0i}+\delta_{i}^{\rho}\delta_{j}^{\mu}w^{[ij]}\\
E^{\rho\mu i} & = & tB^{\rho\mu i}-x^{i}C^{\rho\mu}\,.
\end{eqnarray}

With this choice $w^{\mu i}$ does not enter the boost current 
\begin{equation}
b_{\mathrm{imp}}^{\mu i}=tT_{\mathrm{imp}}^{\mu i}-x^{i}J_{\mathrm{imp}}^{\mu}\,.
\end{equation}
The conservation of $b_{\mathrm{imp}}^{\mu i}$ now implies
\begin{equation}
T_{\mathrm{imp}}^{0i}=J_{\mathrm{imp}}^{i}\,,\label{eq:constraint_Bargmann_current}
\end{equation}
which is to say that the momentum density is the same as the mass flux, a fact known from the general Ward identities of field theories coupled to TNC backgrounds \cite{Jensen:2014aia,Hartong:2014pma}.
We also conclude that the minimal set of independent currents for a generic Bargmann invariant theory is given by $E^\mu,\,T_{\mathrm{imp}}^{\mu i},\,J_\mathrm{imp}^{\mu}$.

\subsection{The coupling of $M_{\mu}$ revisited} \label{Bargmann_noether}

We have shown in section \ref{The minimal Galilean connection} that $\overline{M}_{\mu}$ couples to a topological current $\Phi^{\mu}$ in Galilean theories.
We can learn more about $\Phi^{\mu}$ by examining
the relationship to theories with Bargmann symmetries. Again we will
use the Noether procedure starting with some globally invariant field
theory with action $S^{\left(0\right)}\left[\varphi\right]$. The
extra $U(1)$ symmetry generated by the central charge $N$
 introduces an extra coupling in the $S^{\left(1\right)}$ action involving
a new gauge field $\overline{M}_{\mu}$ that couples to the mass current $J_{\mathrm{can}}^{\mu}$
given by \eqref{Galilei_conserved_currents-1}
\begin{subequations}
\begin{equation}
S\left[\varphi,\overline{e},\overline{\tau},\overline{\Omega}_i,\overline{\Omega}_{ij},\overline{M}\right]= S^{\left(0\right)}\left[\varphi\right]+S^{\left(1\right)}\left[\varphi,\overline{e},\overline{\tau},\overline{\Omega}_i,\overline{\Omega}_{ij},\overline{M}\right]\,,
\end{equation}
where
\begin{equation}
S^{\left(1\right)}=\int_{M}\mathrm{d}^{D}x\,\left[\overline{\tau}_{\mu}E_{\mathrm{can}}^{\mu}+\overline{e}_{\mu i}T_{\mathrm{can}}^{i\mu}-\overline{M}_{\mu}J_{\mathrm{can}}^{\mu}-\overline{\Omega}_{\mu i}w^{\mu i}+\frac{1}{2}\overline{\Omega}_{\mu ij}s^{\mu ij}\right]\,.\label{eq:gauge_couple_bargmann}
\end{equation}
\end{subequations}
We assign to $\overline{M}_{\mu}$ the first order transformation
\begin{equation}
\delta^{\left(1\right)}\overline{M}_{\mu}=-\partial_{\mu}\sigma-\delta_{\mu}^{i}\lambda_{i}\,,\label{eq:transform_gaugefield_Bargmann_noether}
\end{equation}
and the remaining fields transform in the same way as in the Galilean case \eqref{eq:transformations_linear_fields}.

Using the improvements of the boost and mass currents discussed in the previous subsection we can express the canonical
currents in terms of the improved ones. Comparing to the Galilean case we only need to study the new improvements of
the mass current. Performing integration by
parts on these extra terms, we find that we can define new objects that couple to the internal boost and spin current as
\begin{subequations}\label{eq:NoetherBarg}
\begin{equation}
S^{\left(1\right)}=\int_{M}\mathrm{d}^{D}x\,\biggl[\overline{\tau}_{\mu}E_{\mathrm{imp}}^{\mu}+\overline{e}_{0i}T_{\mathrm{imp}}^{i0}+\frac{1}{2}s_{ij}T_{\mathrm{imp}}^{ij}-\overline{M}_{\mu}J_{\mathrm{imp}}^{\mu}+\frac{1}{2}{\overline{Y}}_{\mu ij}s^{\mu ij}-{\overline{Y}}_{\mu i}w^{\mu i}\biggr]\,,
\end{equation}
where we have
\begin{eqnarray}
{\overline{Y}}_{\mu i} & = & \overline{\Omega}_{\mu i}+\delta_{\mu}^{k}\left(\frac{1}{2}\partial_{0}s_{ki}+\partial_{(k}\overline{v}_{i)}\right)+2\delta_{\mu}^{0}\partial_{[0}\overline{M}_{i]}+\delta_{\mu}^{j}\partial_{[j}\overline{M}_{i]}\\
{\overline{Y}}_{\mu ij} & = & \overline{\Omega}_{\mu ij}+\delta_{\mu}^{0}\left(\partial_{0}\overline{e}_{[ij]}+\partial_{[i}\overline{v}_{j]}\right)-\delta_{\mu}^{k}\left(\partial_{[i}s_{j]k}-\partial_{k}\overline{e}_{[ij]}\right)-\delta_{\mu}^{0}\partial_{[i}\overline{M}_{j]}\,.
\end{eqnarray}
\end{subequations}

The connection gauge fields $\overline{\Omega}_{\mu i}\,,\overline{\Omega}_{\mu ij}$ we obtain from this
procedure are identical to the minimal Galilean connection \eqref{noether_pseudo_conn} if we make the choice ${\overline{Y}}_{\mu i}=0$ and ${\overline{Y}}_{\mu ij}=0$.
$S^{(1)}$ is then similar to \eqref{eq:gauge_couple_imprtwo} with no coupling to the internal boost and spin currents,
but now with $\overline{M}_{\mu}$ coupling to $J_{\mathrm{imp}}$.
The origin of $\overline{M}_{\mu}$ is however  different in the two cases.
In the Bargmann case it is a natural gauge field introduced by the Noether procedure, while in the Galilean case it is introduced ad hoc.

As their transformation laws and couplings are the same we can regard the fields
in the two constructions as being identical, which gives a natural interpretation of $\overline{M}_{\mu}$ in Galilean theories: It is the would-be mass gauge field, with the mass current vanishing in these theories. This interpretation is strengthened by  noticing
that the topological current $\Phi^{\mu}$ is exactly of the same form as the
improvements of the mass current we have chosen above.  Hence a unified way of writing the coupling of $\overline{M}_{\mu}$, in Galilean and Bargmann
theories, is to write the current it couples to as
\begin{equation}
J_{\mathrm{imp}}^{\mu}=J_{\mathrm{can}}^{\mu}+\Phi^{\mu}\,.
\end{equation}
Galilean theories have a vanishing canonical mass current $J_{\mathrm{can}}^{\mu}$
and in this case $J_{\mathrm{imp}}^{\mu}$ is pure improvement.

\section{Comparison to the non-linear theory} \label{sec:Comparison to usual procedure}
\subsection{Uniqueness of the minimal connection at non-linear level}

The transformation properties of the gauge fields introduced in the Noether procedure of sections \ref{Galilean_noether} and \ref{Bargmann_noether}
were chosen so that the action is invariant under local Galilean transformations at lowest order.
By comparison they are seen to be identical to the linearized versions of (torsional) Newton-Cartan geometry that we work out in appendix \ref{sec:lin_NC}.
This shows that the Noether procedure, as expected, produces TNC geometry at the linear level. It is therefore appropriate to identify the various gauge fields of the Noether procedure with those of Newton-Cartan geometry as our use of notation suggests.

The connections that emerged from the analysis of the improvement terms are readily identified with those of TNC geometry.
We see that the Galilean/Bargmann connections $\tilde{\Omega}_{ \mu i},\,\tilde{\Omega}_{ \mu ij}$ as given in \eqref{noether_pseudo_conn} are identical to the linearized version of the connections \eqref{eq:connectionsaa} with the choice \eqref{eq:minconnho}. This leads to the affine connection \eqref{gravphotonic_conn} which was proven in \cite{Jensen:2014aia,Hartong:2014pma,Bergshoeff:2014uea} to be the only one linear in $M_\mu$.
We conclude that \eqref{gravphotonic_conn} is the minimal choice of connection for TNC.

\subsection{Other relevant connections with $M_\mu$}

Interestingly enough there is an entire class of affine connections that are constructed from only the vielbeins and $M_\mu$. To construct more general affine connections, one notices that 
\begin{equation}
\tilde{\Phi} = -v^{\mu}M_{\mu}+\frac{1}{2}h^{\mu\nu}M_{\mu}M_{\nu}\,,
\end{equation}
is invariant under local Galilean boosts. This field can be used to construct a more general class of affine connections that is no longer linear in $M_\mu$ and that can be constructed from the objects $\hat{v}^\mu,\,\overline{h}_{\mu\nu}$ and $\tilde{\Phi}$ \cite{Hartong:2015zia} (see also \cite{Bekaert:2014bwa} for a discussion of general TNC connections)
\begin{subequations} 
\begin{eqnarray}
\Gamma_{\mu\nu}^{\lambda} & = & -\hat{v}^{\lambda}\partial_{\mu}\tau_{\nu}+\frac{1}{2}h^{\lambda\sigma}\left(\partial_{\mu}H_{\nu\sigma}\left(\alpha\right)+\partial_{\nu}H_{\mu\sigma}\left(\alpha\right)-\partial_{\sigma}H_{\mu\nu}\left(\alpha\right)\right)\,,\\
H_{\mu\nu}\left(\alpha\right) & = & \overline{h}_{\mu\nu}+\alpha\tilde{\Phi}\tau_{\mu}\tau_{\nu}\,.
\end{eqnarray}
\end{subequations}
Choosing $\alpha = 0$ gives the minimal connection \eqref{gravphotonic_conn}.
The $\alpha \neq 0$ connections are evidently not identified as minimal Galilean connections by the Noether procedure, even though they contain the same number of fields as \eqref{gravphotonic_conn}. This is because for these connections $M_\mu$ doesn't couple only to the mass or topological current, which was an assumption in the Noether procedure.

Another example of a non-minimal connection is given in \cite{Hartong:2015xda}
\begin{equation}\label{eq:manifestinvGamma2TNC}
\check\Gamma^\lambda_{\mu\rho}=-\hat v^\lambda\partial_\mu\tau_\rho+\frac{1}{2}h^{\nu\lambda}\left(\partial_\mu\hat h_{\rho\nu}+\partial_\rho\hat h_{\mu\nu}-\partial_\nu\hat h_{\mu\rho}\right)+\frac{1}{2}h^{\nu\lambda}\tau_\rho \mathcal{L}_{\hat v}\hat h_{\mu\nu}\,,
\end{equation}
where $\hat h_{\mu\nu}=\bar h_{\mu\nu}+2\tilde\Phi\tau_\mu\tau_\nu$ which is such that $\hat v^\mu \hat h_{\mu\nu}=0$. This latter connection has the nice property that $\check\nabla_\mu\tau_\nu=0$, $\check\nabla_\mu\hat v^\nu=0$, $\check\nabla_\mu h^{\nu\rho}=0$ and $\check\nabla_{\mu}\hat h_{\nu\rho}=0$. All of these more general connections correspond to different choices of the contortion-like objects ${\overline{Y}}_{\mu i}$ and ${\overline{Y}}_{\mu ij}$ in \eqref{eq:NoetherBarg}.

\section{Galilean electrodynamics}\label{Galdelectroynamics}

In this section we will discuss an explicit example of a massless Galilean theory, namely Galilean electrodynamics, to illustrate
the general results presented before. Our discussion of the theory will be succinct, and we refer to the reader to the
 the companion paper \cite{Festuccia:2016caf} where we study various versions of non-relativistic electrodynamics and their couplings to TNC geometry in detail. 

\subsection{Action and equations of motion} 

There are a number of ways to obtain non-relativistic theories starting from
known relativistic ones. One may simply take the non-relativistic limit $c\rightarrow\infty$, rescaling the fields properly with powers of $c$,
while another way is to perform null reductions on Lorentz invariant theories.
It has been known for some time \cite{LeBellac:1973} that Maxwellian electrodynamics allows for two non-relativistic limits,
called  the \textit{electric} and \textit{magnetic} limits, respectively.
It is possible to embed both of these limits into a larger theory, that we will call Galilean electrodynamics (GED) and to which we restrict
for simplicity.
This theory consists of three fields $a_i,\,\varphi,\,\tilde{\varphi}$ that are all needed in order to describe dynamical non-relativistic electrodynamics. The fields furnish an indecomposable (but not irreducible) vector representation of the homogeneous Galilean group.
$a_{i}$ here transforms as a ($D-1)$-vector under rotations and mixes with $\varphi$ under boosts, while $\varphi$ is a spacetime
scalar. $\tilde{\varphi}$  mixes with the two other fields under boosts (see below).

The covariant action of GED coupled to TNC is given by \cite{Festuccia:2016caf}
\begin{equation}\label{eq:GED}
S_{\text{GED}}=\int d^{d+1}x e\left(-\frac{1}{4}h^{\mu\rho}h^{\nu\sigma}F_{\mu\nu}F_{\rho\sigma}-h^{\mu\nu}\hat v^\rho F_{\rho\nu}\partial_\mu\varphi-\tilde\Phi h^{\mu\nu}\partial_\mu\varphi\partial_\nu\varphi+\frac{1}{2}\left(\hat v^\mu\partial_\mu\varphi\right)^2\right)\,,
\end{equation}
where $F_{\mu\nu}=\partial_\mu A_\nu-\partial_\nu A_\mu$ with $A_\mu=a_\mu-\tilde\varphi\tau_\mu-\varphi M_\mu$  expressed in terms of the  GED fields  $a_\mu$, $\tilde \varphi$, $ \varphi$  (with $v^\mu a_\mu =0$).  
 $a_\mu$, $\tilde \varphi$ are the gauge potentials entering the magnetic field and electric field, and 
 transform under local Galilean boosts and $U(1)_\Lambda$ gauge transformations as 
\begin{equation}
\label{eq:barcompt}
a_\mu\sim a_\mu + \tau_\mu v^\nu \partial_\nu \Lambda\,,\quad \delta  a_\mu=\varphi\, e_{\mu}^a \lambda_a +\tau_\mu a_\nu e^\nu_a \lambda^a\,,\quad \tilde \varphi\sim \tilde \varphi+v^\nu \partial_\nu\Lambda \,,\quad  \delta \tilde\varphi=a_\nu e^\nu_a \lambda^a\,.
\end{equation}
The field $  \varphi$  is an extra scalar field, called the mass potential, which is invariant under these transformations.
It follows that  $A_\mu$ is  inert under local Galilean boosts and consequently the  action \eqref{eq:GED} is manifestly boost invariant. The action
is also invariant under $U(1)_\sigma$ transformations with  $\delta M_\mu = \partial_\mu \sigma$,   $\delta A_\mu=-\varphi\partial_\mu\sigma$.  

The equations of motion obtained by varying the GED action \eqref{eq:GED} take the form
\begin{equation}
\label{GED:EOM}
\partial_\mu\left(e\,\tilde F^{\mu\nu}\right)=0\,,\qquad\partial_\mu\left(e\, \tilde G^\mu\right)=0\,
\end{equation}
where $\tilde F^{\mu\nu}$ and $\tilde G^\mu$ are defined as
\begin{eqnarray}
\label{eq:eomgedtnc}
\tilde F^{\mu\nu} & = & h^{\mu\rho}h^{\nu\sigma}F_{\rho\sigma}+\left(\hat v^\mu h^{\nu\rho}-\hat v^\nu h^{\mu\rho}\right)\partial_\rho\varphi\,,\\
\tilde G^\mu & = & h^{\mu\nu}\hat v^\rho F_{\rho\nu}+2\hat\Phi h^{\mu\nu}\partial_\nu\varphi-\hat v^\mu\hat v^\nu\partial_\nu\varphi\,.
\end{eqnarray}
and one can verify hat the set of equations \eqref{GED:EOM} are invariant under both $U(1)$ transformations. 
We note that an efficient way of obtaining the action of GED on a general TNC background is via a null reduction of Maxwellian electromagnetism in one dimension higher \cite{Festuccia:2016caf}.

We will also need the form of the action on flat TNC spacetime, for which 
 $\tau_\mu=\delta_\mu^t$, $~e^\mu_a=\delta^\mu_a$, $~v^\mu=-\delta^\mu_t$ and $e^a_\mu=\delta^a_\mu$. The flat space GED fields are given by $a_i=a_\mu e^\mu_i\,,~\tilde\varphi$, and $\varphi$. 
The GED action  \eqref{eq:GED} then reduces to 
\begin{equation}\label{eq:flatGED}
S=\int d^{d+1}x\left(-\frac{1}{4}f^{ij}f_{ij}-\tilde E^i\partial_i\varphi+\frac{1}{2}\left(\partial_t\varphi\right)^2\right)\,.
\end{equation}
where $f_{ij} = \partial_i a_j - \partial_j a_i $ and 
  $\tilde E_i=-\partial_i\tilde\varphi-\partial_ta_i$.
 This action was first introduced in \cite{Santos:2004pq} and the limit from which it arises is described in \cite{Bergshoeff:2015sic}~.
In particular, it can be obtained from an appropriate non-relativistic limit of Maxwell theory on flat space coupled to a massless free
scalar field.

\subsection{Noether currents and linearization of the action} 

GED is a massless Galilean field theory and we now proceed by presenting the results of Sec.~\ref{noether_gal} for this specific theory
and show that we recover the linearized version of the GED action \eqref{eq:GED} on a general TNC background 

\subsubsection*{Canonical Noether currents}

We first use the  results of section \ref{conserved_noether_gal} to determine the canonical Noether currents and
internal boost and spin currents. The calculations are straightforward given the transformation laws of the fields 
(see  \cite{Festuccia:2016caf} for details) and the results are
\begin{subequations}
\begin{eqnarray}
E_{\mathrm{can}}^{0} & = & \frac{1}{2}\partial_{t}\varphi \partial_{t}\varphi -\partial^{i}\tilde{\varphi}\partial_{i}\varphi +\frac{1}{2}\left(\partial^{j}a^{k}\partial_{j}a_{k}-\partial^{j}a^{k}\partial_{k}a_{j}\right) \\
E_{\mathrm{can}}^{j} & = & \partial^{j}\varphi \partial_{t}\tilde{\varphi}+\partial^{j}\tilde{\varphi}\partial_{t}\varphi +\partial_{t}a^{j}\partial_{t}\varphi -\partial^{j}a^{k}\partial_{t}a_{k}+\partial^{k}a^{j}\partial_{t}a_{k} \\
\nonumber \\
T_{\mathrm{can}}^{0i} & = & \partial_{t}\varphi \partial^{i}\varphi +\partial_{k}\varphi \partial^{i}a^{k} \\
T_{\mathrm{can}}^{ji} & = & \partial^{j}\varphi \partial^{i}\tilde{\varphi}+\left(\partial^{j}\tilde{\varphi}+\partial_{t}a^{j}\right)\partial^{i}\varphi +\left(\partial^{i}a_{k}\partial^{k}a^{j}-\partial^{i}a_{k}\partial^{j}a^{k}\right) \\
 &  & +\left(\frac{1}{2}\left(\partial^{n}a^{k}\partial_{n}a_{k}-\partial^{n}a^{k}\partial_{k}a_{n}\right)-\left(\partial^{n}\tilde{\varphi}+\partial_{t}a^{n}\right)\partial_{n}\varphi -\frac{1}{2}\left(\partial_{t}\varphi \right)^{2}\right)\delta^{ij} \label{eq:Canonical_currents_4D_GED} \\
\nonumber\\
b_\mathrm{can}^{\mu i} & = & tT^{\mu i}+w^{\mu i} \\
w^{0 i} & = & -\varphi \partial^{i}\varphi  \\
w^{ji} & = & -a^{i}\partial^{j}\varphi +2\varphi \partial^{[j}a^{i]} \\
\nonumber \\
j_{\mathrm{can}}^{\mu ij} & = & x^{i}T_{\mathrm{can}}^{\mu j}-x^{j}T_{\mathrm{can}}^{\mu i}+s^{\mu ij} \\
s^{0ij} & = & 2a^{[j}\partial^{i]}\varphi  \\
s^{kij} & = & 2a^{[i}\partial^{|k|}A^{j]}-2a^{[i}\partial^{j]}a^{k}\,.
\end{eqnarray}
\end{subequations}
The canonical currents are all of the correct form. Notice that the stress tensor is not symmetric, the boost current
contains the non-conserved internal boost current $w^{\mu i}$ and the rotation
current contains the spin-current $s^{\mu ij}$. Moreover these
currents are not  $U(1)_\Lambda$ invariant.

An easy way to derive the expressions above is to perform a null reduction of the $(D+1)$-dimensional canonical Maxwell EM tensor.
One then finds that the energy and momentum currents above are those corresponding to the first $D$ dimensions 
while the extra $(D+1)$-component of the current will be identical zero. The latter component corresponds in general to the mass current
in the reduced theory, which is naturally zero in GED.

\subsubsection*{Improved currents}

Using the improvements that maximally simplify the conserved currents
as discussed in section \ref{gal current improvements}, we find the relevant improved currents to be
\begin{subequations}\label{eq:GED_improved currents}
\begin{eqnarray}
T_{\mathrm{imp}}^{0 i} & = & \partial_{t}\varphi \partial^{i}\varphi +2\partial_{k}\varphi \partial^{[i}a^{k]} \\
T_{\mathrm{imp}}^{ji} & = & 2\partial^{(i}\varphi \partial^{j)}\tilde{\varphi}+2\partial_{t}a^{(i}\partial^{j)}\varphi +2\partial_{k}a^{(i}\partial^{j)}a^{k} \nonumber\\
&& -\partial^{(i}a_{k}\partial^{j)}a^{k}-\partial^{k}a^{(i}\partial_{k}a^{j)}-\mathcal{L}\delta^{ij}\\
\nonumber\\
b_{\mathrm{imp}}^{\mu i} & = & tT_{\mathrm{imp}}^{\mu i}+\psi^{\mu i} \\
\psi^{0 i} & = & w^{0i}=-\varphi \partial^{i}\varphi  \\
\psi^{k i} & = & w^{[ki]}-\frac{1}{2}s^{0ik}=2\varphi \partial^{[k}a^{i]} \\
\nonumber \\
j_{\mathrm{imp}}^{\mu ij} & = & x^{i}T_{\mathrm{imp}}^{\mu j}-x^{j}T_{\mathrm{imp}}^{\mu i}\,.
\end{eqnarray}
\end{subequations}
all of which are gauge invariant.  Note that $T_{\mathrm{imp}}^{ij}$ is manifestly symmetric as it should be.
The non-conserved part given by $\psi^{\mu i}$ cannot be removed 
entirely from the boost current, while the spin current is not present
in the improved rotation current, in accordance with previous results.

The improvement of the energy
current is not determined by the methods of section \ref{gal current improvements},
but requiring it to be gauge invariant leads to a unique current given by
\begin{subequations}\label{eq:GED_flat_conserved_currents2}
\begin{eqnarray}
E_{\mathrm{imp}}^{0} & = & \mathcal{L}-\partial_{0}\varphi\partial_{0}\varphi-\partial^{i}\varphi\left(\partial_{0}a_{i}+\partial_{i}\tilde{\varphi}\right)\\
E_{\mathrm{imp}}^{i} & = & 2\left(\partial_{0}a_{j}+\partial_{j}\tilde{\varphi}\right)\partial^{[i}a^{j]}-\partial_{0}\varphi\left(\partial_{0}a^{i}+\partial^{i}\tilde{\varphi}\right)\,.
\end{eqnarray}
\end{subequations}
Performing null reduction of the $(D+1)$-dimensional improved Maxwell EM tensor will also give the above currents.
One further sees that the improvements of the $(D+1)$-component corresponding to the would-be mass current is identical to the topological current 
$\Phi^\mu$ defined in \eqref{eq:topological_current_noether}, which takes the form
\begin{equation}
\Phi^{\rho}  =  -\left(\partial_{0}\left(\varphi \partial^{j}\varphi \right)-\partial_{i}\left(2\partial^{[i}A^{j]}\varphi \right)\right)\delta_{j}^{\rho}+\partial_{i}\left(\varphi \partial^{i}\varphi \right)\delta_{0}^{\rho}\,.\label{eq:asdas}
\end{equation}

\subsubsection*{Linearization and Noether procedure}

Finally, we wish to show that linearizing the action \eqref{eq:GED} we obtain the same as compared to 
using the Noether procedure of section \ref{Galilean_noether}. 
Using the linearized version of TNC geometry given in appendix \ref{sec:lin_NC}, we find after straightforward but tedious calculations
that the linearized action \eqref{eq:GED} becomes%
\footnote{There is also an extra term
$(\partial^{i}\varphi \partial_{0}a^{j}+\partial^{i}a^{k}\partial_{k}a^{j})\overline{e}_{[ij]}-2a^{j}\partial^{[k}a^{i]}\partial_{k}\overline{e}_{[ij]}+a^{j}\partial^{i}\varphi \partial_{0}\overline{e}_{[ij]}$ involving terms that are antisymmetric in the vielbein, $\propto\overline{e}_{[ij]}$, which vanishes upon using the flat space GED equations of motion.}
of the form \eqref{eq:gauge_couple_imprtwo} with the improved currents as in \eqref{eq:GED_improved currents}, \eqref{eq:GED_flat_conserved_currents2}
and the topological current given by \eqref{eq:asdas}.

\section{Discussion}

With the Noether procedure it is easier to answer questions about how field theories couple to geometry.
For Bargmann theories it was well-known from the gauging procedure of section \ref{TNC_gauge} that $M_\mu$ would have to couple to a (covariantly) conserved mass current \cite{Jensen:2014aia,Hartong:2014pma}. What was not clear from previous work was the role of $M_\mu$ in massless Galilean theories. The Noether procedure clears this coupling up completely. When $M_{\mu}$ does not couple to the mass current, it remains a background field that couples to the topological current $\Phi^{\rho}$ which is written in terms of the non-conserved spin- and internal boost currents. 

In the companion paper \cite{Festuccia:2016caf} we confirm these results for Galilean electrodynamics, which together with its matter couplings is analyzed extensively there. We in particular work out the coupling to TNC geometry beyond the linearized level studied here. As shown in this paper, when we linearize these couplings we find perfect agreement with the
general results of the Noether method performed here.

An important bonus of the Noether procedure provides is that it makes a natural proposal for what we call the minimal affine connection that comes the closest to being an analog of the Levi-Civita connection in (T)NC geometry. It comes about by improving the canonical Noether currents such that the Galilean boost and rotation Ward identities are made manifest, i.e. the fact the momentum flux equals the mass flux and that the stress tensor is symmetric. If we then only couple the TNC variables $\tau_\mu$, $e^i_\mu$ and $M_\mu$ to these energy, momentum and mass currents, we obtain a form of the affine connection \eqref{gravphotonic_conn} that we refer to as the minimal connection whose non-linear version is the unique TNC connection that is linear in $M_\mu$.

It would be interesting to extend these ideas and calculations to theories with broken Galilean symmetries such as Lifshitz theories or to theories with Carrollian boost invariance%
\footnote{We note that  in two dimensions, Carrollian and Galilean boost invariance are dual to each other under interchange of time and space. 
An example of a 2D theory with this boost symmetry is warped CFT, which was shown in Ref.~\cite{Hofman:2014loa} to couple to
warped geometry. The latter can be seen either as TNC geometry or as Carrollian geometry, which are thus dual to each other in two dimensions
\cite{Hartong:2015xda}.}
 and to see if the Noether procedure leads to the Carrollian geometry of \cite{Bekaert:2015xua,Hartong:2015xda}. Further it would be interesting to see if the Noether procedure could be helpful for the construction of non-relativistic supersymmetric gravity theories \cite{Andringa:2013zja,Bergshoeff:2015uaa,Bergshoeff:2015ija,Bergshoeff:2016lwr}.

In \cite{Hartong:2015zia} it has been shown that dynamical (TT)NC geometry corresponds to (non-) projectable Ho\v rava--Lifshitz gravity
\cite{Horava:2008ih,Blas:2009qj,Horava:2010zj}.  It has been noted that making TTNC geometry dynamical almost always leads to a breaking\footnote{This breaking is described in \cite{Hartong:2015zia} by adding a St\"uckelberg scalar $\chi$ to the theory.} of the $U(1)$ symmetry associated with the Bargmann extension. It is presently not very clear why this is what happens generically. The only known exceptions are in 3 dimensions corresponding the Chern--Simons versions of Ho\v rava--Lifshitz gravity \cite{Hartong:2016yrf}. It would be interesting to see if the Noether procedure can shine some light on this issue either by explaining why this happens or by making clear that there must exist a larger space of Ho\v rava--Lifshitz type theories that does have a $U(1)$ gauge symmetry.

\section*{Acknowledgments}

We would like to thank  Jan de Boer, Diego Hofman and Cynthia Keeler for useful discussions. 
The work of GF is supported by the ERC STG grant 639220. 
The work of JH is partially supported by a Marina Solvay fellowship as well as by the advanced ERC grant `Symmetries and Dualities in Gravity and M-theory' of Marc Henneaux. 
The work of NO is supported in part by the Danish National Research Foundation project 
 ``New horizons in particle and condensed matter physics from black holes". GF is supported by the ERC STG grant 639220.

\appendix

\section{The Noether procedure for Lorentzian theories\label{sec:The-Noether-procedure-lorentz}}

For the convenience of comparison with a more familiar setting we review here the Noether procedure for Lorentzian theories.

\subsection{Conserved Noether currents} \label{currents_lorentz}

It is insightful to see how the Noether procedure works in the more
familiar case of theories with Poincar\'e spacetime symmetries.
The starting point for this is again an analysis of the current multiplet of these theories, (see for instance  \cite{Dumitrescu:2011iu}).
 We assume here a Lagrangian description with action $S^{\left(0\right)}\left[\varphi\right]=\int_{M}\mathrm{d}^{D}x\,\mathcal{L}\left(\varphi,\partial\varphi,x\right)$
and linear transformations given by
\begin{subequations}
\begin{eqnarray}
\delta x^{\mu} & = & \epsilon^{\mu}+\lambda^\mu{}_{\nu} x^{\nu}\\
\delta\varphi_{\ell} & = & \frac{1}{2}\lambda_{\mu\nu}\left(J^{\mu\nu}\right)_{\ell\ell^{\prime}}\varphi_{\ell^{\prime}}\,,
\end{eqnarray}
\end{subequations}
where $\lambda_{\mu\nu}=-\lambda_{\nu\mu}$ is just an infinitesimal
Lorentz transformation and $\epsilon^{\mu}$ an infinitesimal translation
that together are all of the parameters of the transformation. The
generators of rotations $\left(J^{\mu\nu}\right)_{\ell\ell^{\prime}}$
act on the fields $\varphi_{\ell}$ in some suitable linear representation.

The corresponding conserved Noether currents are given by

\begin{subequations}
\begin{eqnarray}
T_{\mathrm{can}}^{\rho\mu} & = & \frac{\partial\mathcal{L}}{\partial\left[\partial_{\rho}\varphi_{\ell}\right]}\partial^{\mu}\varphi_{\ell}-\eta^{\mu\rho}\mathcal{L}\\
j_{\mathrm{can}}^{\rho\mu\nu} & = & x^{\mu}T_{\mathrm{can}}^{\rho\nu}-x^{\nu}T_{\mathrm{can}}^{\rho\mu}+s^{\rho\mu\nu}\\
\nonumber \\
s^{\rho\mu\nu} & = & -\frac{\partial\mathcal{L}}{\partial\left[\partial_{\rho}\varphi_{\ell}\right]}\left(J^{\mu\nu}\right)_{\ell\ell^{\prime}}\varphi_{\ell^{\prime}}
\end{eqnarray}
\end{subequations}

where $T_{\mathrm{can}}^{\rho\mu}$ is the canonical energy-momentum
tensor and $j_{\mathrm{can}}^{\rho\mu\nu}$ is the total angular momentum
containing the non-conserved spin-current $s^{\rho\mu\nu}$. The conservation
$\partial_{\rho}j_{\mathrm{can}}^{\rho\mu\nu}=0$ implies
\begin{equation}
2T^{[\mu\nu]}_{\mathrm{can}}=-\partial_{\rho}s^{\rho\mu\nu}\,,
\end{equation}
so in general $T_{\mathrm{can}}^{\mu\nu}\neq T_{\mathrm{can}}^{\nu\mu}$.

We can write a variation of the action in terms of arbitrary local
parameters $\lambda_{\mu\nu}\left(x\right)$ and $\xi_{\mu}\left(x\right)\equiv\epsilon_{\mu}+\lambda_{\mu\nu}x^{\nu}$
as
\begin{equation}
\delta S^{\left(0\right)}=-\int_{M}\mathrm{d}^{D}x\,\left[(\partial_{\rho}\xi_{\mu}-\lambda_{\mu\rho})T^{\rho\mu}_{\mathrm{can}}+\frac{1}{2}\partial_{\rho}\lambda_{\mu\nu}s^{\rho\mu\nu}\right]\,,\label{eq:poincare_local_variation}
\end{equation}

Global transformations correspond to constant $\epsilon_\mu$ and $\lambda_{\mu\nu}$.

\subsection{Improvements of currents}

The improvements of the EM tensor and angular momentum current take the form
\begin{subequations}\label{eq:Poincare_improved_currents}
\begin{eqnarray}
T_{\mathrm{imp}}^{\rho\mu} & = & T_{\mathrm{can}}^{\rho\mu}+\partial_{\lambda}A^{\lambda\rho\mu}\,,\\
J_{\mathrm{imp}}^{\rho\mu\nu} & = & x^{\mu}T_{\mathrm{can}}^{\rho\nu}-x^{\nu}T_{\mathrm{can}}^{\rho\mu}+s^{\rho\mu\nu}+\partial_{\lambda}B^{\lambda\rho\mu\nu}
\end{eqnarray}
\end{subequations}

where $A^{\lambda\rho\mu}=-A^{\rho\lambda\mu}$ and $B^{\lambda\rho\mu\nu}=-B^{\rho\lambda\mu\nu}$
are the improvement terms. If we choose
\begin{subequations}
\begin{eqnarray}
A^{\lambda\rho\mu} & = & \frac{1}{2}\left(s^{\lambda\rho\mu}+s^{\rho\mu\lambda}-s^{\mu\lambda\rho}\right)\label{eq:Improvement_EM_Poincare}\\
B^{\lambda\rho\mu\nu} & = & x^{\mu}A^{\lambda\rho\nu}-x^{\nu}A^{\lambda\rho\mu}\,,
\end{eqnarray}
\end{subequations}
then the conservation law for angular momentum reads

\begin{equation}
T_{\mathrm{imp}}^{\mu\nu}=T_{\mathrm{imp}}^{\nu\mu}
\end{equation}

This improved energy-momentum tensor is  known as the Belinfante-Rosenfeld energy-momentum tensor.

\subsection{Coupling the gauge fields to currents} \label{Noether_proc_Lorentz}

If we want local Poincar\'e invariance of our theory at lowest order, we need a first-order term in the action that cancels
the non-invariance of (\ref{eq:poincare_local_variation}). We define
two gauge fields that couple to the currents in $S^{\left(1\right)}$
\begin{equation}
S^{\left(1\right)}=\int_{M}\mathrm{d}^{D}x\,\left[\overline{e}_{\mu\nu}T_{\mathrm{can}}^{\mu\nu}+\frac{1}{2}\overline{\omega}_{\rho\mu\nu}s^{\rho\mu\nu}\right]\,.
\end{equation}
These are the linearization of the vielbein $\overline{e}_{\mu\nu}$ and the
spin-connection $\overline{\omega}_{\rho\mu\nu}$ and transform at first order
as
\begin{subequations}
\begin{eqnarray}
\delta^{\left(1\right)}\overline{e}_{\mu\nu} & = & \partial_{\mu}\xi_{\nu}+\lambda_{\mu\nu}\\
\delta^{\left(1\right)}\overline{\omega}_{\rho\mu\nu} & = & \partial_{\rho}\lambda_{\mu\nu}
\end{eqnarray}
\end{subequations}

Writing $T_{\mathrm{can}}^{\nu\mu}$ in terms of the improved current
\eqref{eq:Poincare_improved_currents} gives after some algebra
\begin{subequations}
\begin{equation}
S^{\left(1\right)}=\int_{M}\mathrm{d}^{D}x\,\left[\frac{1}{2}\overline{h}_{\mu\nu}T_{\mathrm{imp}}^{\mu\nu}+\frac{1}{2}\overline{Y}_{\rho\mu\nu}s^{\rho\mu\nu}\right]
\end{equation}
\begin{equation}
\overline{Y}_{\rho\mu\nu}=\overline{\omega}_{\rho\mu\nu}-\overline{\omega}_{(0)\rho\mu\nu}\,,
\end{equation}
\begin{equation}
\overline{\omega}_{(0)\rho\mu\nu}=\frac{1}{2}\left(\partial_{\rho}\overline{h}_{\mu\nu}-\partial_{\nu}\overline{h}_{\mu\rho}\right)+\partial_{\mu}\overline{e}_{[\nu\rho]}\,,
\end{equation}
\end{subequations}

where $\overline{h}_{\mu\nu}=2\overline{e}_{(\mu\nu)}$ is the
perturbation of the Minkowski metric, and one may check that $\overline{\omega}_{(0)\rho\mu\nu}$
is the linearization of the Levi-Civita connection. What one should appreciate is that contrary to the Galilean case,
it is possible to realize the geometry on the vielbeins alone
(or equivalently the metric $g_{\mu\nu}=\eta_{\mu\nu}+\overline{h}_{\mu\nu}+\ldots$).

Minimal coupling  corresponds  to $\overline{Y}_{\rho\mu\nu}=0$. 
In this case, we know from the fundamental theorem of Riemannian geometry that the Levi-Civita connection
is the unique metric compatible and torsionless connection.
This allows to conclude, even at the non-linear level, that the Levi-Civita connection is the minimal connection of Riemannian geometry.
$\overline{Y}_{\rho\mu\nu}$ therefore is interpreted as the (linearized) contortion tensor (which can be expressed in terms of the torsion tensor). Thus minimal coupling to gravity, metric compatibility and torsionlessness of the spin-connection are equivalent for Poincar\'e invariant theories.

\section{Linearization of Newton-Cartan geometry\label{sec:lin_NC}}

The TNC geometry was reviewed in section \ref{TNC_geom}. It simplifies substantially when we only consider it at the linear level, which is sufficient to see the effects of gravity.

We will consider small perturbations to global inertial frames of
the flat geometry considered in \cite{Hartong:2015wxa} and keep everything
at first order. We take
\begin{subequations}\label{eq:linearized_vielbeine}
\begin{eqnarray}
\tau_{\mu} & = & \delta_{\mu}^{0}+\overline{\tau}_{\mu}\\
v^{\mu} & = & -\delta_{0}^{\mu}-\overline{v}^{\mu}\\
e_{\mu}^{i} & = & \delta_{\mu}^{i}+\overline{e}_{\mu}^{i}\\
e_{i}^{\mu} & = & \delta_{i}^{\mu}-\overline{e}_{i}^{\mu}\,,
\end{eqnarray}
\end{subequations}
where $\overline{e}_{\mu}^{i}\,,\overline{e}_{i}^{\mu}\,,\overline{\tau}_{\mu}\,,\overline{v}^{\mu}$
are the perturbations. The completeness relation
must still hold at first order, which implies some relations between
the vielbeins and their inverses:
\begin{subequations}
\begin{eqnarray}
\overline{\tau}_{0} & = & -\overline{v}^{0}\\
\overline{e}_{i}^{0} & = & \overline{\tau}_{i}\\
\overline{e}_{0}^{i} & = & -\overline{v}^{i}\,.
\end{eqnarray}
\end{subequations}
This shows that the linearized inverse vielbeins are proportional to the linearized vielbeins themselves. The spatial metric
can then be expressed in terms of the vielbeins up to first order
as
\begin{subequations}
\begin{eqnarray}
\overline{h}_{\mu\nu} & = & \delta_{ij}\delta_{\mu}^{i}\delta_{\nu}^{j}+\overline{s}_{\mu\nu}=\left(\begin{array}{cc}
0 & -\overline{v}_{j}\\
-\overline{v}_{i} & \delta_{ij}+\overline{s}_{ij}
\end{array}\right)\\
\overline{h}^{\mu\nu} & = & \delta^{ij}\delta_{i}^{\mu}\delta_{j}^{\nu}-\overline{s}^{\mu\nu}=\left(\begin{array}{cc}
0 & -\overline{\tau}^{j}\\
-\overline{\tau}^{i} & \delta^{ij}-\overline{s}^{ij}
\end{array}\right)\,,
\end{eqnarray}
\end{subequations}
where we have defined the perturbations of the spatial metrics as
\begin{subequations}
\begin{eqnarray}
\overline{s}_{\mu\nu} & = & \overline{e}_{\mu}^{i}\delta_{\nu i}+\delta_{\mu}^{i}\overline{e}_{\nu i}=2\delta_{(\mu}^{i}\overline{e}_{\nu)i}\\
\overline{s}^{\mu\nu} & = & \overline{e}_{i}^{\mu}\delta^{\nu i}+\delta_{i}^{\mu}\overline{e}^{\nu i}=2\delta_{i}^{(\mu}\overline{e}^{\nu)i}\,.
\end{eqnarray}
\end{subequations}
All spatial indices may be raised or lowered with the flat spatial metric $\delta^{ij}\,,\delta_{ij}$.

With these results we can linearize the affine pseudo-connection \eqref{eq:pseudo_connection} and the corresponding gauge fields. The results are
\begin{subequations}
\begin{equation}
\overline{\Gamma}_{(0)\mu\nu}^{\lambda}=\delta_{0}^{\lambda}\partial_{\mu}\overline{\tau}_{\nu}+\frac{1}{2}\delta^{\lambda i}\delta_{i}^{\sigma}\left(\partial_{\mu}\overline{s}_{\nu\sigma}+\partial_{\nu}\overline{s}_{\mu\sigma}-\partial_{\sigma}\overline{s}_{\mu\nu}\right)\,,\label{eq:pseudo_connection-1}
\end{equation}
\begin{eqnarray}
\overline{\Omega}_{(0) \mu i} & = & \left(\begin{array}{c}
\overline{\Omega}_{(0) 0 i}\\
\overline{\Omega}_{(0) j i}
\end{array}\right)=\left(\begin{array}{c}
0\\
-\frac{1}{2}\partial_{0}\overline{s}_{ji}-\partial_{(i}\overline{v}_{j)}
\end{array}\right)\\
\overline{\Omega}_{(0) \mu ik} & = & \left(\begin{array}{c}
\overline{\Omega}_{(0) 0 ik}\\
\overline{\Omega}_{(0) j ik}
\end{array}\right)=\left(\begin{array}{c}
-\partial_{0}\overline{e}_{[ik]}-\partial_{[i}\overline{v}_{k]}\\
-\partial_{j}\overline{e}_{[ik]}+\partial_{[i}\overline{s}_{k]j}
\end{array}\right)\,.
\end{eqnarray}
\end{subequations}
The full gauge fields can then be written as
\begin{subequations}
\begin{eqnarray}
\overline{\Omega}_{\mu i} & = & \overline{\Omega}_{(0) \mu i}+\overline{C}_{\mu i}\\
\overline{\Omega}_{\mu ij} & = & \overline{\Omega}_{(0) \mu ij}+\overline{C}_{\mu ij}\,,\\
\nonumber\\
\overline{C}_{\mu i} & = & \delta_{\lambda i}\overline{W}^\lambda{}_{\mu0}\\
\overline{C}_{\mu ij} & = & \delta_{\lambda i}\overline{W}^\lambda{}_{\mu j}\,.
\end{eqnarray}
\end{subequations}

The relevant transformation laws \eqref{NC_transformations_infs} are
\begin{subequations}\label{eq:Linearized_Gal_transform}
\begin{eqnarray}
\delta\overline{\tau}_{\mu} & = & \partial_{\mu}\xi^{0}\,,\\
\delta\overline{e}_{\mu}^{i} & = & \partial_{\mu}\xi^{i}+\lambda^i{}_{j}\delta_{\mu}^{j}-\lambda^{i}\delta_{\mu}^{0}\\
\delta\overline{\Omega}_{\mu}{}_{i} & = & -\partial_{\mu}\lambda_{i}\label{eq:first_order_gauge_Galilean}\\
\delta\overline{\Omega}_{\mu}{}_{ij} & = & \partial_{\mu}\lambda_{ij}\,.
\end{eqnarray}
\end{subequations}
For the minimal connection \eqref{gravphotonic_conn} we take the mass/background gauge field $M_{\mu}=\overline{M}_{\mu}$ and give it the transformation law
\begin{equation}
\delta\overline{M}_{\mu}=-\partial_{\mu}\sigma-\lambda^{i}e_{\mu i}\,.
\end{equation}
We then find that the pseudo-contortions become
\begin{subequations}
\begin{eqnarray}
\overline{C}_{\mu i} & = & -\left(\begin{array}{c}
2\partial_{[0}\overline{M}_{i]}\\
\partial_{[j}\overline{M}_{i]}
\end{array}\right)\\
\overline{C}_{\mu ij} & = & \left(\begin{array}{c}
\partial_{[i}\overline{M}_{j]}\\
0
\end{array}\right)\,.
\end{eqnarray}
\end{subequations}

\addcontentsline{toc}{section}{\refname}

\providecommand{\href}[2]{#2}\begingroup\raggedright\endgroup


\end{document}